\documentclass[12pt]{article}

\usepackage{slashed}
\usepackage{latexsym}
\newcommand{\SU}{\mathop{\rm SU}}
\newcommand{\SO}{\mathop{\rm SO}}
\newcommand{\UU}{\mathop{\rm {}U}}

\newcommand{\Red}[1]{#1}
\newcommand{\Blue}[1]{#1}
\def\be{\begin{equation}}
\def\ee{\end{equation}}
\def\bea{\begin{eqnarray}}
\def\eea{\end{eqnarray}}
\def\nn{\nonumber}
\def\qq{\quad\quad}

\newcommand{\ft}[2]{{\textstyle\frac{#1}{#2}}}

\def\del{\partial}

\def\a{\alpha}
\def\b{\beta}
\def\g{\gamma}

\def\d{\delta}

\def\e{\epsilon}
\def\ve{\varepsilon}

\def\f{\phi}

\def\p{\psi}

\def\vf{\varphi}

\def\l{\lambda}
\def\L{\Lambda}
\def\m{\mu}
\def\n{\nu}
\def\r{\rho}
\def\s{\sigma}

\def\th{\theta}

\def\o{\omega}

\def\g{\gamma}
\def\x{\xi}
\def\rmi{{\,\rm i\,}}

\csname @addtoreset\endcsname{equation}{section}

\begin{document}

\begin{titlepage}
\begin{flushright}
  UG-01-01 \\
    KUL-TF-01/11 \\
  hep-th/0104113
\end{flushright}

\begin{center}
\vspace{.5cm} \baselineskip=16pt {\LARGE \bf Weyl multiplets of \\[3mm] $N=2$
conformal supergravity\\[3mm]  in five dimensions } \vskip 0.8 cm {\large
  Eric Bergshoeff$^1$, Sorin Cucu$^2$,\\[3mm] Martijn Derix$^2$, Tim de
Wit$^1$,\\[3mm]
  Rein Halbersma$^1$ and Antoine Van Proeyen$^2$
} \\
\vskip 8 mm
{\small
  $^1$ Institute for Theoretical Physics, University of Groningen\\
       Nijenborgh 4, 9747 AG Groningen, The Netherlands\\
       \{e.bergshoeff, t.c.de.wit, r.halbersma\}@phys.rug.nl \\[3mm]
  $^2$ Instituut voor Theoretische Fysica, Katholieke Universiteit Leuven\\
       Celestijnenlaan 200D B-3001 Leuven, Belgium\\
       \{sorin.cucu, martijn.derix, antoine.vanproeyen\}@fys.kuleuven.ac.be
}
\end{center}

\centerline{ABSTRACT}
\bigskip

We construct the Weyl multiplets of $N=2$ conformal supergravity in five
 dimensions. We show that there exist two different versions of the Weyl
multiplet, which contain the same gauge fields but differ in the matter
field content: the Standard Weyl multiplet and the Dilaton Weyl multiplet.
At the linearized level we obtain the transformation rules for the Dilaton
Weyl multiplet by coupling it to the multiplet of currents corresponding
to an on-shell vector multiplet. We construct the full non-linear
transformation rules for both multiplets by gauging the $D=5$
superconformal algebra $F^2(4)$. We show that the Dilaton Weyl multiplet
can also be obtained by
 solving the equations of motion for an improved vector multiplet coupled
to the Standard Weyl multiplet.

\end{titlepage}

\tableofcontents

%%%%%
\section{Introduction}

Conformal supergravities have been constructed in various dimensions (for
a review, see~\cite{Salam:1989fm}) but not yet in five dimensions. The
five-dimensional case is of interest for various reasons not least of all
from a purely mathematical viewpoint since it is based on the exceptional
superalgebra $F^2(4)$.

By using conformal tensor calculus, conformal supergravities form an
elegant way to construct general couplings of Poincar{\'e}-supergravities
to matter~\cite{Kaku:1978ea}. In the five-dimensional case these matter
coupled supergravities have recently attracted renewed attention due to
the important role they play in the Randall--Sundrum (RS)
scenario~\cite{Randall:1999vf,Randall:1999ee} and the
$AdS_6/CFT_5$~\cite{Nishimura:2000wj,D'Auria:2000ad} and
$AdS_5/CFT_4$~\cite{Balasubramanian:2000pq} correspondences.

The form of the scalar potential in five-dimensional matter coupled
supergravities plays a crucial role in the possible supersymmetrisation
of the RS-scenario. It turns out that such a supersymmetrisation is
non-trivial. With only vector multiplets and no singular source
insertions, a no-go theorem was established  for smooth domain-wall
solutions~\cite{Kallosh:2000tj,Behrndt:2000tr}. In view of this, general
$D=5$ supergravity/matter couplings have been
re-investigated~\cite{Ceresole:2000jd}, thereby generalizing the earlier
results of~\cite{Gunaydin:1984bi,Gunaydin:1985ak}. A modification of the
theory allows solutions by inserting branes as singular
insertions~\cite{Bergshoeff:2000zn}. The inclusion of hypermultiplets was
first considered in~\cite{Behrndt:2000km}, where even generalizations
of~\cite{Ceresole:2000jd} were considered. However, this description has
not been proven to be consistent. Hypermultiplets were also considered
in~\cite{Behrndt:2000ph,Behrndt:2001qa}. The mixing of vector and
hypermultiplets~\cite{Ceresole:2001wi} seems to circumvent all
obstructions, though no example of a good smooth solution has been found.
However, it has been shown also in~\cite{Ceresole:2001wi} that $N=2$,
$D=5$ matter couplings to supergravity can give rise to more general
possibilities for renormalization group flows between conformal theories
in ultraviolet and infrared than those known for $N=8$.

With all these developments, it is clear that it is important that there
is an independent derivation of the most general matter couplings derived
in~\cite{Ceresole:2000jd}. Moreover, it has turned out in the past that
superconformal constructions lead to insights in the structure of matter
couplings. A recent example is the insight in relations between
hyper-K{\"a}hler cones and quaternionic manifolds, based on the study of
superconformal invariant matter couplings with
hypermultiplets~\cite{deWit:2001dj}. For all these reasons a
superconformal construction of general matter couplings in $N=2$, $D=5$
is useful.

In this paper we take the first step in this investigation by
constructing the $N=2$, $D=5$ conformal supergravity theory. In our
construction we use the methods developed first for $N=1$,
$D=4$~\cite{Kaku:1977pa,Kaku:1978nz}. They are based on gauging the
conformal superalgebra~\cite{Ferrara:1977ij} which in our case is
$F^2(4)$.

The superconformal multiplet that contains all the (independent) gauge
fields of the superconformal algebra is called the Weyl multiplet. In
general one needs to include matter fields to have an equal number of
bosons and fermions. We will see that in five dimensions there are two
possible sets of matter fields one can add, yielding two versions of the
Weyl multiplet: the Standard Weyl multiplet and the Dilaton Weyl
multiplet. This result is similar to what was found for $(1,0)\ D=6$
conformal supergravity theory~\cite{Bergshoeff:1986mz}. Also in that case,
two versions were found: a multiplet containing a dilaton and one without
a dilaton.

In~\cite{Nishimura:2000wj}, the field content and transformation rules for
the Standard Weyl multiplet were constructed from the $F(4)$-gauged
six-dimensional supergravity using the $AdS_6/CFT_5$ correspondence. The
results, although not given in a manifestly superconformal notation, seem
similar to ours. However, the full non-linear commutation relations
(see~(\ref{algebraQQ})) that we obtain were not given.

Another attempt was undertaken in~\cite{Kugo:2000hn} by reducing the known
six-dimensional result~\cite{Bergshoeff:1986mz} to five dimensions. The
authors of~\cite{Kugo:2000hn} already gauge-fixed some symmetries of the
superconformal algebra during the reduction process in order to simplify
the matter multiplet coupling. In this way, they found a multiplet that
is larger than the Weyl multiplet that we will construct in this work,
because they do not aim to obtain superconformal symmetry in 5
dimensions. Our strategy is to start from the basic building blocks of
superconformal symmetry in 5 dimensions.

We first construct the conformal supercurrent multiplet that contains the
energy--momentum tensor of the $D=5$ vector multiplet. This is non-trivial
because the $D=5$ vector multiplet is \emph{not} conformal. At first sight
this seems to prohibit the construction of a \emph{conformal} current
multiplet, but we will show how the introduction of a dilaton in the Weyl
multiplet circumvents this obstacle. This is the origin of the first of
our two versions of 32+32 Weyl multiplets. The other one is a
straightforward extension of the one known in 4 dimensions.

We have organized the paper such that a reader who is interested in the
main results for the multiplets, i.e.\ their content, transformation laws
and the algebra that they satisfy can find everything in
section~\ref{ss:finalWeyls}. The rules found in this section will be
needed when one investigates matter couplings. However, this does not
contain all our results. The relation between the two versions is based
on the use of the (improved) vector multiplet, and this construction is
also part of our main result.

This paper is organized as follows. In section~\ref{ss:linWeyl}, as the
first step in our procedure, we construct the supercurrent multiplet that
contains the energy-momentum tensor of the $N=2$, $D=5$ vector multiplet.
It turns out that this supercurrent multiplet has $32+32$ components.

The coupling of the supercurrent multiplet to the fields of conformal
supergravity leads to the linearized superconformal transformation rules
for the $32 + 32$ component Dilaton Weyl multiplet. We show that there
exists another version of the linearized Weyl multiplet (the Standard Weyl
multiplet) that contains the same gauge fields as the Dilaton Weyl
multiplet, but differs in the matter field content. An important
difference between the Standard and Dilaton Weyl multiplet is that the
scalar field of the Standard Weyl multiplet has a non-zero mass dimension
that cannot serve, like the dilaton scalar field of the Dilaton Weyl
multiplet, as a compensator for scale transformations.

In section~\ref{ss:gaugingSC} we derive the full non-linear transformation
rules for both Weyl multiplets by gauging the $D=5$ superconformal algebra
$F^2(4)$ following the notations on real forms as
in~\cite{VanProeyen:1999ni}. For the convenience of the reader we give
the final results of the two Weyl multiplets, in a self-contained manner,
in section~\ref{ss:finalWeyls}.

In section~\ref{ss:connection} we show that the Dilaton Weyl multiplet
can be obtained by coupling the Standard Weyl multiplet to an improved
vector multiplet. This establishes the precise connection between the two
multiplets.  We present our conclusions in section~\ref{ss:conclusions}.

We explain our notation and conventions in appendix~\ref{app:conventions}.
The complete commutation relations defining the $D=5$ superconformal
algebra $F^2(4)$ are given in appendix~\ref{app:algF4}. Finally, in
appendix~\ref{ss:howlin} we compare the $32+32$ supercurrent multiplet we
construct in this paper with the $40+40$ supercurrent multiplet
constructed by Howe and Lindstr{\"o}m~\cite{Howe:1981nz} some time ago. We
show that their multiplet is reducible.

\section{Linearized Weyl multiplets}\label{ss:linWeyl}

In this section we obtain two linearized Weyl multiplets. After discussing
the method of the supercurrent (section~\ref{ss:currentmethod}) we will
construct the currents of a rigid on-shell vector multiplet
(section~\ref{ss:currMult}), and define a Weyl multiplet as the fields
that couple to the currents (section~\ref{ss:lindilW}). The comparison
with known Weyl multiplets in 4 and 6 dimensions, tells us that there is
also another Weyl multiplet, and we point out that it can be obtained
from the first one by redefining some fields (section~\ref{ss:linstW}).

\subsection{The current multiplet method} \label{ss:currentmethod}

The multiplet of currents in a superconformal context has been discussed
before in the literature, e.g.~the current multiplet corresponding to
 the $N=1$,
$D=4$~\cite{Kaku:1978nz}, the $N=2$,
$D=4$~\cite{Ferrara:1975pz,Sohnius:1979pk} and the $N=4$, $D=4$ vector
multiplets~\cite{Bergshoeff:1981is} and to the (self-dual) $(2,0)\ D=6$
tensor multiplet~\cite{Bergshoeff:1999db}.

After adding local improvement terms one obtains a supercurrent multiplet
containing an energy-momentum tensor $\th_{\m\n} = \th_{\n\m}$
and a supercurrent
 $J^i_\m$ which are both conserved and (gamma-)traceless
\be
\partial^\m \th_{\m\n} = \th_\m^{~\m} = \partial^\m J^i_\m = \g^\m J^i_\m = 0\,.
 \ee
These improved current multiplets were used in the past to construct the
linearized transformation rules for the Weyl multiplet\footnote{The Weyl
multiplets of $(1,0)\ D=6$~\cite{Bergshoeff:1986mz} were derived without
the use of a current multiplet, although this is certainly possible in
view of the reduction rules given in~\cite{Bergshoeff:1999db}.} since a
traceless energy-momentum tensor is equivalent to scale-invariance of the
kinetic terms in the action.

However, the standard kinetic term of the $D=5$ vector field
\begin{equation}
\label{kinetic}
{\cal L} = - \ft14 F_{\mu\nu}F^{\mu\nu}
\end{equation}
is not scale invariant, i.e.\ the energy-momentum tensor is not traceless:
\begin{equation}
\th_{\m\n} = - F_{\m\l} F_\n^{~\l} + \ft14 \eta_{\m\n} F_{\rho\sigma}
F^{\rho\sigma}\, ,\hskip 2truecm \theta_\mu{}^\mu = \ft14 F_{\m\n} F^{\m\n}
\ne 0\, .
\end{equation}
Moreover, there do not exist gauge-invariant local improvement terms.

There is a remedy for this problem. Whenever there is a compensating
scalar field present, i.e.~a scalar with mass dimension zero but non-zero
Weyl weight, then the kinetic term~(\ref{kinetic}) can be made scale
invariant by introducing a scalar coupling of the form
\begin{equation}
{\cal L} = - \ft14 e^\phi F_{\mu\nu}F^{\mu\nu}\, .
\end{equation}
This compensating scalar is called the dilaton. In general, there are
three possible origins for a dilaton coupling to a non-conformal matter
multiplet: the dilaton is part of
\begin{enumerate}
  \item the matter multiplet itself (the multiplet is then called an `improved'
  multiplet);
  \item the conformal supergravity multiplet;
  \item another matter multiplet.
\end{enumerate}
The $N=2$, $D=5$ vector multiplet contains precisely such a scalar. We
could therefore use it to compensate the broken scale invariance of the
kinetic terms. This leads to the so-called improved vector multiplet. This
is the first possibility, that will be further discussed in
section~\ref{ss:connection}.

The second possibility will be considered here (the third possibility is
included for completeness). This possibility thus occurs when the Weyl
multiplet itself contains a dilaton. We will see that there indeed exists
a version of the Weyl multiplet containing a dilaton. This version is
called the Dilaton Weyl multiplet. It turns out that there exists another
version of the Weyl multiplet without a dilaton. This other version will
be called the Standard Weyl multiplet.

For matter multiplets having a traceless energy-momentum tensor, no
compensating scalar is needed. To see the difference between the various
cases it is instructive to consider $(1,0)\ D=6$  conformal supergravity
theory~\cite{Bergshoeff:1986mz} which was constructed without the
supercurrent method. In that case, two versions were found: a multiplet
containing a dilaton and one without a dilaton. We expect that both
versions can be constructed using the supercurrent method: the one
without a dilaton starting from the conformal $(1,0)$ tensor multiplet
(being a truncation of the $(2,0)$ case), and the version containing the
dilaton by starting from the non-conformal $D=6$ vector multiplet (which
upon reduction should produce our results in $D=5$).

Thus, the current multiplet needs to be improved only when coupled to the
Standard Weyl multiplet. In the case of the Dilaton Weyl multiplet it is
not necessary to do so, since in that case the dilaton of the Weyl
multiplet can be used to compensate for the lack of scale invariance. In
particular, the dilaton will couple directly to the trace of the
energy-momentum tensor.

When coupling to the Standard Weyl multiplet one needs to add
\emph{non-local} improvement terms to the current multiplet which was
done for the current multiplet coming from the $D=10$ vector
multiplet~\cite{Bergshoeff:1982av}. In that case the non-local
improvement terms that were added, required the use of auxiliary fields
satisfying differential constraints in order to make the transformation
rules local.\footnote{Note also that in $D=10$ the trace-part and the
traceless part of the energy-momentum tensor are not contained in the same
multiplet which necessitates the addition of the non-local improvement
terms to project out the trace-part.}

We did not analyse the addition of non-local counter terms. It would be
interesting to see if in this way a consistent coupling to the Standard
Weyl multiplet can be obtained. Instead, we will derive the linearized
transformation rules for the Standard Weyl multiplet via a field
redefinition from those of the Dilaton Weyl multiplet.

\subsection{Current multiplet of the $N=2$, $D=5$ vector
multiplet}\label{ss:currMult}

Our starting point is the on-shell $D=5$ vector multiplet. Its field
content is given by a massless vector $A_\m$, a symplectic Majorana
spinor $\p^i$ in the fundamental of $\SU(2)$ and a real scalar $\s$. See
table~\ref{tbl:nAvm}
\begin{table}[htb]
\begin{center}
\begin{tabular}{||c|c|c|c|c||}
\hline
\rule[-1mm]{0mm}{6mm}
Field       & Equation of motion   & {$\SU(2)$} & $w$ & \# d.o.f.  \\
\hline
\rule[-1mm]{0mm}{6mm}
$A_\m$      & $\partial_\m F^{\m\n} = 0$ & 1 & 0             & 3 \\
\rule[-1mm]{0mm}{6mm}
$\s$        & $\Box \s = 0$          & 1 & 1             & 1 \\
\hline \rule[-1mm]{0mm}{6mm} $\p^i$      & $\slashed{\partial} \p^i =
0$   & 2 & ${3/ 2}$ &
4 \\[1mm]
\hline
\end{tabular}
\caption{\it The $4+4$ on-shell abelian vector multiplet.\label{tbl:nAvm}}
\end{center}
\end{table}
for additional information. Our conventions are given in
appendix~\ref{app:conventions}.

The action for the $D=5$ Maxwell multiplet is given by
 \be {\cal L} = -\ft14 F_{\m\n}F^{\m\n} - \ft12 \bar{\p} \slashed{\partial}
\p -
\ft12 (\partial \s)^2\,. \label{eq:onshell_maxwell}
 \ee
 This action is invariant under the following supersymmetries
 \bea
\d_Q A_\m &=& \ft12 \bar{\e} \g_\m \p\, , \nonumber\\
\d_Q \p^i &=& -\ft14 \g \cdot F \e^i - \ft12\rmi \slashed{\partial}
\s \e^i\, , \nonumber\\
\d_Q \s   &=& \ft12\rmi \bar{\e} \p\, ,
 \eea
  as well as under the standard gauge transformation
 \be \d_\Lambda  A_\m = \partial_\m \L\, . \ee

The various symmetries of the lagrangian~(\ref{eq:onshell_maxwell}) lead
to a number of Noether currents: the energy-momentum tensor
$\theta_{\mu\nu}$, the supercurrent $J_\mu^i$ and the $\SU(2)$-current
$v_\mu^{ij}$. The  supersymmetry variations of these currents lead to a
closed multiplet of $32+32$ degrees of freedom (see
table~\ref{tbl:currentmult}). As discussed in the introduction, an
unconventional feature, compared to the currents corresponding to a $D=4$
vector multiplet or a $D=6$ tensor multiplet, is that the current
multiplet cannot be improved by local gauge-invariant terms,
i.e.~$\theta_{\mu}{}^\mu \ne 0$ and $\gamma^\mu J_\mu^i \ne 0$. It is
convenient to include these trace parts as separate currents since, as it
turns out, they couple to independent fields of the Weyl multiplet.

\begin{table}[htb]
\begin{center}
\begin{tabular}{||c|c|c|c|c|c|c||}
\hline
\rule[-1mm]{0mm}{6mm}
Current        & Noether & {$\SU(2)$} & $w$ &
\# d.of. \\
\hline
\rule[-1mm]{0mm}{6mm}
$\th_{(\m\n)}$ & $\partial^\m \th_{\m\n} = 0$  & 1 & 2              & 9 \\
\rule[-1mm]{0mm}{6mm}
$\th_{\m}{}^{\m}$       &  & 1 & 4              & 1 \\
\rule[-1mm]{0mm}{6mm}
$v^{(ij)}_\m$  & $\partial^\m v^{ij}_\m =0$    & 3 & 2              & 12 \\
\rule[-1mm]{0mm}{6mm}
$a_\m$       & $\partial^\m a_\m = 0$       & 1 &  3             & 4 \\
\rule[-1mm]{0mm}{6mm}
$b_{[\m\n]}$   & $\partial^\m b_{\m\n} = 0$   & 1 &  2             & 6 \\
\hline
\rule[-1mm]{0mm}{6mm}
$J^i_\m$     & $\partial^\m J^i_\m = 0$     & 2 & ${5 / 2}$  & 24  \\
\rule[-1mm]{0mm}{6mm}
$\zeta^i \equiv \rmi \g \cdot J^i$     & & 2 & ${7 / 2}$  & 8 \\[1mm]
\hline
\end{tabular}
\caption{\it The $32+32$ current multiplet. The trace  $\theta_\mu {}^\mu
$ and the gamma-trace of $J^i_\mu $ form separate currents, the latter is
denoted by $\zeta^i$.
}\label{tbl:currentmult}
\end{center}
\end{table}

We find the following expressions for the Noether currents and their
supersymmetric partners in terms of bilinears of the vector multiplet
fields:
 \bea \th_{\m\n} &=& - \partial_\m \s \partial_\n \s +\ft12
\eta_{\m\n} \left(\partial \s\right)^2 -  F_{\m\l} F_\n^{~\l} +
\ft14 \eta_{\m\n} F^2 -
\ft12\bar{\p}  \g_{(\mu }\partial_{\nu )} \p\, , \nonumber\\
J^i_\m &=& -\ft14 \rmi \g \cdot F \g_\m \p^i - \ft12 (\slashed{\partial}
\s) \g_\m \p^i\, , \nonumber\\
v^{ij}_{\m} &=& \ft12 \bar{\p}^i \g_{\m} \p^j\, , \nonumber\\
 a_\m
&=& \ft18 \ve_{\m\n\l\r\s} F^{\n\l} F^{\r\s} + (\partial^\n \s)
F_{\n\m}\, , \nonumber\\
 b_{\m\n}
&=&\ft12  \ve_{\m\n\l\r\s} (\partial^\l \s) F^{\r\s} +\ft12
 \bar{\p}  \g_{[\mu }\partial_{\nu ]} \p\, , \nonumber\\
\zeta^i &=& \rmi \g \cdot J^i = \ft14 \g \cdot F \p^i +  \ft32 \rmi
\slashed{\partial} \s \p^i\, ,
 \nonumber\\
\th_\m^{~\m} &=& \ft32 \left(\partial \s\right)^2 + \ft14 F^2\, . \eea
{}From these expressions, using the Bianchi identities and equations of
motion of the vector multiplet fields, one can calculate the
supersymmetry transformations of the currents. A straightforward
calculation yields:
 \bea \d_Q \th_{\m\n}
&=& \ft12\rmi \bar{\e} \g_{\l(\m}\partial^\l  J_{\n)}\, , \nonumber\\
\d_Q J^i_\m &=& - \ft12 \rmi \g^\n \th_{\m\n} \e^i - \rmi
\g_{[\l}\partial^\l v^{ij}_{\m]} \e_j - \ft12 a_\m \e^i + \ft12\rmi
\g^\n b_{\m\n} \e^i\, , \nonumber\\
\d_Q v^{ij}_\m
&=& \rmi \bar{\e}^{(i} J^{j)}_\m\, , \nonumber\\
\d_Q a_\m &=& -\bar{\e} \partial^\l \g_{[\l} J_{\m]} + \ft14 \bar{\e}
\g_{\r\m} \g^\s \partial^\r J_\s +  \ft14\rmi \bar{\e} \g_{\r\m}
\partial^\r \zeta\, , \nonumber\\
\d_Q b_{\m\n} &=& \ft34\rmi \bar{\e}  \g_{[\l\m}\partial^\l J_{\n]} -
\ft18\rmi \bar{\e} \g_{\r\m\n} \g^\l \partial^\r J_\l + \ft18 \bar{\e}
\g_{\r\m\n} \partial^\r \zeta\, , \nonumber\\
\d_Q \zeta^i &=& \ft12 \rho \e^i - \ft12 \slashed{\partial}
\slashed{v}^{ij} \e_j -\ft12 \rmi \slashed{a} \e^i -
\ft12 \g \cdot b \e^i\, , \nonumber\\
\d_Q \th_\m^{~\m} &=& \ft12 \bar{\e} \slashed{\partial} \zeta\, .
\label{lintrcurr} \eea

Note that we have added to the transformation rules for $a_\mu$ and
$b_{\mu\nu}$ terms that are identically zero: the first term at the
r.h.s.\ contains the divergence of the supercurrent and the last two terms
are proportional to the combination $(i \g \cdot J - \zeta)$ which is
zero. Similarly, the second term in the variation of the supercurrent
contains a term that is proportional to the divergence of the $\SU(2)$
current.

The reason why we added these terms is that in this way we obtain below
the linearized Weyl multiplet in a conventional form. Alternatively, we
could not have added these terms and later have brought the Weyl
multiplet into the same conventional form by redefining the
$Q$-transformations via a field-dependent $S$- and $\SU(2)$-transformation.

\subsection{Linearized Dilaton Weyl multiplet} \label{ss:lindilW}

The linearized $Q$-supersymmetry transformations of the Weyl multiplet are
determined by coupling every current to a field, and demanding invariance
of the corresponding action. The field-current action is given by:
 \be
\label{eq:cur2lin}
 S = \int d^5 x \bigg(\ft12 h_{\m\n} \th^{\m\n} + \rmi
\bar{\p}_\m J^\m + V_\m^{ij}v_{ij}^\m + A_{\m} a^{\m} + B_{\m\n} b^{\m\n}
+ \rmi \bar{\p} \zeta + \vf\ \th_\m^{~\m} \bigg)\, . \ee In
table~\ref{tbl:fieldsWeyls} we give some properties of the Weyl
multiplets. In particular of the one just derived, which we call the
Dilaton Weyl multiplet\footnote{Note that the Dilaton Weyl multiplet
contains a vector $A_\mu$, a spinor $\psi^i$ and a scalar $\sigma$ which,
on purpose, we have given the same names as the fields of the vector
multiplet. The reason for doing so will become clear in
section~\ref{ss:connection} where we explain the connection between the
two Weyl multiplets. From now on, until section~\ref{ss:connection}, we
will be only dealing with the Weyl multiplets and not with the vector
multiplet. Therefore, our notation should not lead to confusion.}. A
similar Weyl multiplet containing a dilaton exists in
$D=6$~\cite{Bergshoeff:1986mz}.
\bigskip

\begin{table}[ht]
\begin{center}
$  \begin{array}{||c|cccc||c|cccc||} \hline\hline
\mbox{Field} & \#  & \mbox{Gauge}&\SU(2)& w & \mbox{Field} & \#  &
\mbox{Gauge}&\SU(2)& w\\
\hline \rule[-1mm]{0mm}{6mm}& \multicolumn{4}{c||}{\mbox{Elementary gauge
fields}} && \multicolumn{4}{c||}{\mbox{Dependent gauge fields}}\\
\rule[-1mm]{0mm}{6mm} \Blue{e_\mu{} ^a & 9 & P^a &1&-1 & \omega _\mu
^{[ab]}&- & M^{[ab]}}&1&0   \\ \rule[-1mm]{0mm}{6mm}
\Blue{b_\mu   & 0 & D &1&0&f_\mu{} ^a &-& K^a &1&1 }  \\
\rule[-1mm]{0mm}{6mm}
\Blue{V_\mu ^{(ij)} &12  & \SU(2)\,&3&0&& & &  & }  \\
\hline \rule[-1mm]{0mm}{6mm}
\Red{\psi _\mu ^i  & 24 & Q^i_\alpha
&2&-1/2 & \phi _\mu ^i & - &
 S^i_\alpha&2&1/2} \\ [1mm]
\hline\hline \rule[-1mm]{0mm}{6mm} & \multicolumn{4}{c||}{\mbox{Dilaton
Weyl multiplet}} && \multicolumn{4}{c||}{\mbox{Standard Weyl
multiplet}}\\ \rule[-1mm]{0mm}{6mm} \Blue{A_\mu  & 4 & \d A_\m =
\partial_\m \L &1&0 & T_{[ab]}&  10& & 1&1 } \\ \rule[-1mm]{0mm}{6mm}
\Blue{B_{[\mu \nu] }  & 6 & \d B_{\m\n} = 2 \partial_{[\m}
\L_{\n]}\hspace{-2pt} &1&0 &&& &  & } \\
\rule[-1mm]{0mm}{6mm} \Blue{\varphi & 1 & &1&1 & D&1 &  & 1&2 }\\ \hline
\rule[-1mm]{0mm}{6mm} \Red{\p ^i  & 8 & &2&3/2 & \chi ^i & 8& & 2&3/2
}\\ [1mm]
\hline\hline
\end{array}$
  \caption{\it Fields of the Weyl multiplets, and their roles. The upper
 half contains the fields
  that are present in all versions. They are the gauge fields of the
superconformal algebra (see section ~\ref{ss:gaugingSC}). The fields at
the right-hand side of the upper half are dependent fields, and are not
visible in the linearized theories. The symbol $\#$ indicates the
off-shell degrees of freedom. The gauge degrees of freedom corresponding
to the gauge invariances of the right half are subtracted from the fields
at the left on the same row. In the lower half are the extra matter
fields that appear in the two versions of the Weyl multiplet. In the left
half are those of the Dilaton Weyl multiplet, at the right are those of
the Standard Weyl multiplet. We also indicated the (generalized) gauge
symmetries of the fields $A_\m$ and $B_{\m\n}$. (The linearized fields,
corresponding to $e_\mu{}^a$ and $\s \equiv e^\f$ are denoted by
$h_\mu{}^a$ and $\vf$, respectively.) }\label{tbl:fieldsWeyls}
\end{center}
\end{table}

Using the supersymmetry rules for the current multiplet, we find that the
following transformations leave the action~(\ref{eq:cur2lin}) invariant:
 \bea \d_Q h_{\m\n}
&=& \bar{\e} \g_{(\m} \p_{\n)}\, , \nonumber\\
\d_Q \p^i_\m &=& - \ft14 \g^{\l\n} \partial_\l h_{\n\m} \e^i - V_\m^{ij}
\e_j  + \ft18\rmi \bigg(\g \cdot F + \ft13\rmi \g \cdot H\bigg)
\g_\m \e^i\, , \nonumber\\
\d_Q V^{ij}_\m &=& - \ft12 \bar{\e}^{(i} \g^\l \p^{j)}_{\l\m} \nn
 + \ft12\rmi \bar{\e}^{(i} \g_\m \slashed{\partial} \p^{j)}\, , \nonumber\\
\d_Q A_\m &=& -\ft12\rmi \bar{\e} \p_\m + \ft12 \bar{\e} \g_\m \p\, ,
\nonumber\\
\d_Q B_{\m\n} &=& \ft12 \bar{\e} \g_{[\m} \p_{\n]} + \ft12\rmi \bar{\e}
\g_{\m\n} \p\, , \nonumber\\
 \d_Q \p^i &=& -\ft18 \g \cdot F \e^i -\ft12\rmi \slashed{\partial}
\vf \e^i + \ft1{24} \rmi \g \cdot H \e^i\, ,
\nonumber\\
\d_Q \vf &=& \ft12 \rmi \bar{\e} \p\, , \label{lintrDilW}
 \eea
 where we have defined
\be F_{\m\n} = 2 \partial_{[\m} A_{\n]}, \quad H_{\m\n\l} = 3
\partial_{[\m} B_{\n\l]}, \quad \p_{\m\n} = 2 \partial_{[\m} \p_{\n]}\, .
\ee

\subsection{Linearized Standard Weyl multiplet}\label{ss:linstW}

It turns out that there exists a second formulation of the Weyl multiplet
in which the fields $A_\m$ and $B_{\m\n}$ are replaced by an
anti-symmetric tensor $T_{ab}$ and where also the spinor and the scalar
are redefined. It is the multiplet we should have expected if we compare
it with the Weyl multiplets of the $D=4$ and $D=6$ theories with 8
supercharges. This can be seen in table~\ref{tbl:countWeyl}.
\begin{table}[ht]
\begin{center}
$ \begin{array}{|l|rrr|} \hline\hline
 \mbox{Field}  & d=4 & d=5 & d=6 \\
\hline\hline
e_\mu {}^a     & 5 & 9 & 14 \\
b_\mu    & 0 &0  &0 \\
\omega _\mu {}^{ab}&- &- & - \\
f_\mu {}^a &- &-&- \\
V_{\mu i}{}^j   & 9 & 12 & 15  \\
A_\mu &3 & - & - \\ \hline &&&\\[-4mm]
\psi _\mu {}^i  & 16 & 24 & 32 \\[1mm]
\phi _\mu {}^i &- &- &- \\[1mm] \hline\hline
T_{ab}, T_{abc}^-  & 6 & 10 & 10 \\
D          & 1 &  1 &  1\\
\hline &&&\\[-4mm]
\chi ^i  & 8 & 8 & 8  \\[1mm]
\hline\hline
\mbox{TOTAL}  & 24+24 & 32+32 & 40+40\\
\hline\hline
\end{array}$
\end{center}
  \caption{\it Number of components in the fields of the Standard Weyl
multiplet.
  The dependent fields have no number. The field $T$ is a two rank tensor in 4
  dimensions and a self-dual three rank tensor in 6 dimensions. In 5
  dimensions we can choose between a two-rank or a three-rank tensor as
  these are dual to each other.
}
  \label{tbl:countWeyl}
\end{table}

This second Weyl multiplet is called the Standard Weyl multiplet. More
information about the component fields can be found in
table~\ref{tbl:fieldsWeyls}. The Standard Weyl multiplet cannot be
obtained  from the same current multiplet procedure we applied to get the
Dilaton Weyl multiplet, unless we would consider an `improved' current
multiplet. The reason is that the Standard Weyl multiplet contains no
dilaton scalar with a zero mass dimension that can be used as a
compensating scalar. Therefore it can not define a conformal coupling to
a non-improved current multiplet.

In 5 dimensions the full superconformal algebra cannot be realized on a
matter multiplet without `improvement' by a dilaton. In
section~\ref{ss:connection} we will explain how the two Weyl multiplets
can be related to each other via the coupling of the Standard Weyl
multiplet to an improved vector multiplet.

The connection between the two versions of the Weyl multiplet at the
linearized level is given by algebraic relations. First of all we denote
some particular terms in the transformations of $\p^i_\m$ and $V^{ij}_\m$ by
 $T_{ab}$ and $\chi^i$. Then we compute the variations of these expressions
 under supersymmetry, finding one more object called $D$. We find
 \bea
T_{ab}
&=& \ft18 \bigg(F_{ab} - \ft16 \ve_{abcde} H^{edc} \bigg)\, ,\nonumber\\
\chi^i
&=& \ft18\rmi \slashed{\partial} \p^i + {1 \over 64} \g^{ab} \p_{ab}\, ,
\nonumber\\
D &=& \ft14 \Box \vf - \ft1{32} \partial^\m \partial^\n h_{\m\n} +
\ft1{32} \Box h^\m_\m\, .\label{linrelW}
 \eea
The resulting supersymmetry transformations are those of what we call the
linearized Standard Weyl multiplet. They
are given by
 \bea \d_Q h_{\m\n}
&=& \bar{\e} \g_{(\m} \p_{\n)}\, , \nonumber\\
\d_Q \p^i_\m &=& - \ft14 \g^{\l\n} \partial_\l h_{\n\m} \e^i - V_\m^{ij}
\e_j + \rmi \g \cdot T \g_\m \e^i\, , \nonumber\\
 \d_Q V^{ij}_\m &=& -\ft18
\bar{\e}^{(i} \bigg(\g^{ab} \g_\m - \ft12 \g_\m \g^{ab}\bigg) \p^{j)}_{ab}
 + 4 \bar{\e}^{(i} \g_\m \chi^{j)}\, , \nonumber\\
\d_Q T_{ab} &=& \ft12\rmi \bar{\e} \g_{ab} \chi -\ft3{32}\rmi \bar{\e}
\bigg(\p_{ab} -\ft1{12} \g_{ab} \g^{cd} \p_{cd} + \ft23 \g_{[a}
\g^c \p_{b]c} \bigg)\, , \nonumber\\
 \d_Q \chi^i &=& \ft14 D \e^i - {1
\over 64} \g^{\m\n} V^{ij}_{\m\n} \e_j + \ft3{32}\rmi \g \cdot T
\overleftarrow{\slashed{\partial}} \e^i + \ft1{32}\rmi \slashed{\partial}
\g \cdot T \e^i\, , \nonumber\\
 \d_Q D &=& \bar{\e} \slashed{\partial}
\chi\, , \label{lintrStW}
 \eea
where we have defined
 \be V^{ij}_{\m\n} = 2 \partial_{[\m} V^{ij}_{\n]}\,. \ee

This concludes our discussion of the linearized Weyl multiplets.

\section{Gauging the superconformal algebra}\label{ss:gaugingSC}

We now proceed with the construction of the full Weyl multiplets, of
which we have shown so far the linearized structure. We apply the methods
developed first for $N=1$ in 4 dimensions~\cite{Kaku:1978nz}. They are
based on gauging the conformal superalgebra~\cite{Ferrara:1977ij}, which,
in our case, is $F^2(4)$. The commutation relations defining the $F^2(4)$
algebra are given in appendix~\ref{app:algF4}. We first discuss the
general method, and then apply this to construct the full (non-linear)
Weyl multiplets for both versions that we found at the linearized level
in section~\ref{ss:linWeyl}. For clarity, we have collected the final
results  in section~\ref{ss:finalWeyls}.
%%%%%
\subsection{The gauge fields and their curvatures}

The $D=5$ conformal supergravity theory is based on the superconformal
algebra $F^2(4)$ whose generators are those in
table~\ref{tbl:superconformal5},
\begin{table}[htb]
\begin{center}
\begin{tabular}{|c|c|c|c|c|c|c|c|}
\hline \rule[-1mm]{0mm}{6mm}
Generators & $P_a$ & $M_{ab} $& $D$ & $K_a$ & $U_{ij}$ & $Q_{\a i}$ &
$S_{\a i}$\\
\hline \rule[-1mm]{0mm}{6mm}
Fields & $e_\m{}^a$ & $\o_\m^{ab}$ & $b_\m$ & $f_\m{}^a$ & $V_\m^{ij}$ &
$\p_\m^i$ & $\phi_\m^i$\\
 \hline
\rule[-1mm]{0mm}{6mm}
Parameters & $\xi^a$ & $\l^{ab}$ & $\L_D$ & $\L_K^a$ & $\L^{ij}$ & ${\e}^i$ &
${\eta}^i$\\
\hline
\end{tabular}
\caption{\it  The gauge fields  and parameters of the superconformal
algebra $F^2(4)$. \label{tbl:superconformal5}}
\end{center}
\end{table}
where $a,b,\dots$ are Lorentz indices, $\a$ is a spinor index and $i=1,2$ is
an $\SU(2)$ index. $M_{ab}$ and $P_a$ are the Poincar{\'e}
generators, $K_a$ is the special conformal transformation, $D$ the
dilatation, $Q_{i\alpha}$ and $S_{i\alpha}$ are the supersymmetry and the
special supersymmetry generators, respectively, which are symplectic
Majorana spinors, 8 real components in total. Finally, $U^{ij} = U^{ji}$
are the $\SU(2)$ generators. For more details on the $F^2(4)$ algebra and
the rigid superconformal transformations, see~\cite{VanProeyen:1999ni}.
The commutation relations of the generators are given in
appendix~\ref{app:algF4}.

As a first step we assign to every generator of the superconformal
algebra a gauge field. These gauge fields and the names of the
corresponding gauge parameters are given in
table~\ref{tbl:superconformal5}.

The transformations are generated by operators according to
 \be \d = \xi^a P_a +\lambda^{ab}M_{ab}
+\Lambda_D D + \Lambda_K^a K^a +\Lambda^{ij} U_{ij} +\rmi \bar\e
Q+\rmi\bar\eta S\, . \label{deltaQeps}
 \ee
The $\rmi$ factors in the last two terms appear due to the reality
properties, as explained in appendix~\ref{app:conventions}.

We can read off the transformation rules for the gauge fields from the
algebra~(\ref{suf2(4)}) using the general rules for gauge theories. We
find
 \bea\label{ftr}
\d e_\m{}^a & = & {\cal D}_\m \x^a -\l^{ab}e_{\m b} -\L_De_\m{}^a +\ft
12{\bar\e} \g^a\p_\m \, , \nn\\
\d \o_\m{}^{ab} & = & {\cal D}_\m\l^{ab} - 4\x^{[a}f_\m{}^{b]} - 4\L_K^{[a}
e_\m{}^{b]}
{+\ft 12}\rmi{\bar\e} \g^{ab} \f_\m
{-\ft 12}\rmi{\bar\eta} \g^{ab} \p_\m\, , \nn \\
\d b_\m & = & \partial_\m \L_D -2 \x^a f_{\m a} +2\L_K^a e_{\m a}
{+\ft 12}\rmi {\bar\e} \f_\m + {\ft 12}\rmi {\bar\eta} \p_\m\, , \nn\\
\d f_\m{}^a & = & {\cal D}_\m \L_K^a - \l^{ab}
f_{\m b} + \L_D f_\m{}^a {+\ft 12}{\bar\eta} \g^a \f_\m\, ,\nn\\
\d V_\m^{ij} & = & \partial_\m\L^{ij} -2\L^{(i}{}_\ell V_\m^{j)\ell} -{\ft
{3}2} \rmi{\bar\e}^{(i}\f_\m^{j)} +{\ft {3}2}\rmi{\bar\eta}^{(i}\p_\m^{j)}
  \,,\nn\\
\d \p_\m^i & = & {\cal D}_\m \e^i {+\rmi}\x^a\g_a\f_\m^i -\ft14 \l^{ab}
\g_{ab}\p_\m^i -\ft12\L_D\p_\m^i
-\L^i{}_j\p_\m^j
{-\rmi}e_\m^a\g_a\eta^i \,,\\
\d \f_\m^i & = & {\cal D}_\m \eta^i -\ft14 \l^{ab}\g_{ab}\f_\m^i +\ft12\L_D
\f_\m^i
-\L^i{}_j\f_\m^j
-{\rmi}\L_K^a\g_a\psi_\m^i +{\rmi}f_\m^a\g_a\e^i  \nonumber \,,
 \eea
 where ${\cal D}_\m$ is the covariant derivative with respect to
dilatations, Lorentz rotations and $\SU(2)$ transformations:
 \bea
{\cal D}_\m \x^a & = & \partial_\m \x^a + b_\m \x^a + \o_\m{}^{ab} \x_b \, ,
\nn \\
{\cal D}_\m \l^{ab} & = & \partial_\m \l^{ab} +2\o_{\m c}{}^{[a} \l^{b]c}\, ,
\nn\\
{\cal D}_\m \L_K^a & = & \partial_\m \L_K^a -b_\m \L_K^a +\o_\m{}^{ab}\L_{Kb}
\, , \nn \\
{\cal D}_\m \e^i & = & \partial_\m \e^i+\ft 12 b_\m\e^i +\ft 14 \o_\m^{ab}
\g_{ab}\e^i - V^{ij}_\m\epsilon _j\,,\nn\\
{\cal D}_\m \eta^i & = & \partial_\m \eta^i-\ft 12 b_\m\eta^i +\ft 14
\o_\m^{ab}\g_{ab}\eta^i - V^{ij}_\m\eta_j\,. \label{covder2}
 \eea
 Using the commutator expressions~(\ref{suf2(4)}) we obtain the following
expressions for the curvatures (terms proportional to vielbeins are
underlined  for later use):
 \bea\label{fcurv} R^{~~~a}_{\m\n}(P) & = &
2\del_{[\m} e_{\n]}{}^a + \underline{2\o_{[\m}{}^{ab} e_{\n]b}} +
\underline{2b_{[\m} e_{\n]}{}^a } {-\ft 12}{\bar\p}_{[\m}\g^a\p_{\n ]}\,,
\nn \\
R_{\m\n}{}^{ab}(M) & = & 2\del_{[\m} \o_{\n]}{}^{ab} + 2\o_{[\m}{}^{ac}
\o_{\n]c}{}^b + \underline{8 f_{[\m}{}^{[a} e_{\n]}{}^{b]}}
+\rmi{\bar\f}_{[\m}\g^{ab} \p_{\n ]}
\, , \nn \\
R_{\m\n}(D) & = & 2\del_{[\m} b_{\n]} -\underline{ 4 f_{[\m}{}^a e_{\n]a}}
{-\rmi}{\bar\f}_{[\m}\p_{\n ]}\, , \nn \\
R_{\m\n}^{~~~a}(K) & = & 2\del_{[\m} f_{\n]}{}^a +2 \o_{[\m}{}^{ab} f_{\n]b} -
2b_{[\m} f_{\n]}{}^a {-\ft 12}{\bar\f}_{[\m}\g^a\f_{\n ]}\,,\nn\\
R_{\m\n}{}^{ij}(V) & = & 2\del_{[\m} V_{\n]}{}^{ij} -2V_{[\m}{}^{k( i}
V_{\n ]\,k}{}^{j)}
{-3\rmi}{\bar\f}^{( i}_{[\m}\p^{j)}_{\n ]} \, ,\\
R_{\m\n}{}^i(Q) & = & 2\del_{[\m}\p^i_{\n ]} +\ft12
\o_{[\m}{}^{ab} \g_{ab}\p^i_{\n ]} +b_{[\m}\p^i_{\n ]}
-2V_{[\m}{}^{ij}\p_{\n ]\,j}
+ \underline{2\rmi\g_a\f^i_{[\mu }e_{\nu ]}{}^a}\,,\nn\\
R_{\m\n}{}^i(S) & = & 2\del_{[\m}\f^i_{\n ]} +\ft12
\o_{[\m}{}^{ab} \g_{ab}\f^i_{\n ]} -b_{[\m}\f^i_{\n ]} -  2V_{[\m}{}^{ij}
\f_{\n ]\,j}
-2\rmi\g_a\p^i_{[\mu}f_{\nu]}{}^a\,. \nonumber
 \eea

Since the transformation laws given above satisfy the $F^2(4)$
superalgebra, we have a gauge theory of $F^2(4)$, but we do not have a
gauge theory of diffeomorphisms of spacetime. This can only be realized
if we take the spin connection as a composite field that depends on the
vielbein. So far, we have it as an independent field.\footnote{One might
think that the field equations can determine the spin connection as a
dependent gauge field.
 This can indeed be done for the spin connection, but it is not known how to
 generalize
this for the gauge fields of special (super)conformal symmetries, which we
also want to be dependent gauge fields.}

Furthermore, we see that the number of
bosonic and fermionic degrees of freedom do not match. The gauge fields
together have $96 + 64$ degrees of freedom. Therefore, we can not have a
supersymmetric theory with invertible general coordinate transformations
generated by the square of supersymmetry operations.

%%%%%%
\subsection{Constraints and their solutions}

The solution to the problems described above is well known. In order to
convert the $P$-gauge transformations into general coordinate transformations
 and to
obtain irreducibility we need to impose curvature constraints and we have
to introduce extra matter fields in the multiplet.

The constraints will
define some gauge fields as dependent fields. The extra matter fields will also
change the transformations of the gauge fields. In fact, we will have for
the transformation (apart from the general coordinate transformations) of
a general gauge field $h_\mu ^I$:
\begin{equation}
  \delta _J(\epsilon ^J)h_\mu ^I= \partial _\mu \epsilon ^I
  +\epsilon ^J h_\mu{}^Af_{AJ}{}^I+\epsilon ^JM_{\mu J}{} ^I\,,
 \label{transfogauge}
\end{equation}
where we use the index $I$ to denote all gauge transformations apart from
general coordinate transformations, and an index $A$ includes the
translations.

The last term depends on the matter fields, and its explicit
form has to be determined below. But also the second term has
contributions from matter fields. This is due to the fact that the structure
`functions' of the final algebra $f_{IJ}{}^K$ are modified from those of
the $F^2(4)$ algebra which was used for~(\ref{ftr}). These extra terms
lead also to modified curvatures
\begin{equation}
  \widehat{R}_{\mu \nu }{}^I=2\partial _{[\mu }h_{\nu ]}{}^I
  +h_\nu {}^Bh_\mu {}^Af_{AB}{}^I-2h_{[\mu }{}^JM_{\nu ]J}{}^I\,.
 \label{hatR}
\end{equation}

The commutator of two supersymmetry-transformations will also change. In
particular we will find transformations with field-dependent parameters.
They can be conveniently written as so-called covariant general
coordinate transformations which are defined as
\begin{equation}
\delta_{cgct}(\xi)=\delta_{gct}(\xi)-\delta_I(\xi^\mu h_\mu{}^I)\,,
\label{defcgct}
\end{equation}
namely a combination of general coordinate transformations and all the other
transformations whose parameter $\epsilon ^I$ is replaced by $\xi^\mu
h_\mu{}^I$. This takes a simpler form on fields of various types:
\begin{eqnarray}
  \delta _{cgct}(\xi )e_\mu {}^a  & = & (\partial _\mu+b_\mu ) \xi ^a+
\omega _{\mu }{}^{ab}\xi
  _b\,, \nonumber\\
   \delta _{cgct}(\xi )h_\mu ^I   & = & -\xi ^\nu \widehat{R}_{\mu \nu }{}^
I-\xi ^\nu  h_\mu ^JM_{\nu J}{}^I
   -\xi ^\nu  h_\mu ^Je_\nu ^af_{a J}{}^I\,,\nonumber\\
%   \delta _{cgct}(\xi )B_{\mu\nu }   & = & \xi ^\rho  \widehat{H}_{\rho \mu
%\nu } +\xi ^\rho
%   \left(\rmi \s \bar \psi _{[\mu }\gamma _{\nu ]\rho }\psi - \ft14 \s^2 \bar{\psi}_\m \g_\r \psi_\n  \right)
%   \,,\nonumber\\
%&& -\xi ^\rho A_{[\m} \left( F_{\nu ]\rho} - \ft12 \rmi \s \bar{\psi}_\r \psi_{\n]} + \ft12 \bar{\psi}_\r \g_{\n]} \psi \right) \nn\\
     \delta_{cgct}(\xi )\Phi &=&\xi ^\mu D_\mu \Phi\,.
 \label{cgctfields}
\end{eqnarray}
The last terms for $B_{\m\n}$ are similar to the $M$-term for usual gauge
fields in the second line. The last line holds for all covariant matter
fields, including their covariant derivatives $D_a$, or covariant
curvature tensors after changing the indices to local Lorentz indices.

We will consider the f{\"u}nfbein as an invertible field. Then some of the
curvatures in~(\ref{fcurv}) are linear in some gauge fields. This is
shown by the underlined terms in~(\ref{fcurv}). Therefore, we can impose
constraints on these curvatures that are solvable for these gauge fields.
Such constraints are called conventional constraints, and imposing them
reduces the Weyl multiplet, such that we get closer to an irreducible
multiplet. The conventional constraints are\footnote{Note that the third
constraint implies that
$\gamma_{[\mu\nu} \widehat R_{\rho\sigma]}{}^i(Q) = 0$.}
 \bea
\label{constraints}
R_{\mu\nu}{}^a(P) = 0 & (50)\nn\,, \\
e^{\nu}{}_b \widehat R_{\mu\nu}{}^{ab} (M) = 0 & (25)\nn\,,\\
\g^\m \widehat R_{\m\n}{}^i (Q) = 0 & (40)\,.
 \eea

In brackets we denoted the number of restrictions each constraint imposes.
These constraints are similar to those for other Weyl multiplets in 4
dimensions with $N=1$~\cite{Kaku:1977pa,Ferrara:1977ij},
$N=2$~\cite{deWit:1980ug} or $N=4$~\cite{Bergshoeff:1981is}, or in 6
dimensions for the $(1,0)$~\cite{Bergshoeff:1986mz} or
$(2,0)$~\cite{Bergshoeff:1999db} Weyl multiplets.

In general one can add extra terms to the constraints~(\ref{constraints}),
which just amount to redefinitions of the composite fields. By choosing
suitable terms simplifications were obtained in 4 and 6 dimensions. In
this case one could e.g.\ add a term $T_{\m a} T^{ab}$ to the second
constraint rendering all the constraints invariant under
$S$-supersymmetry, but in 5 dimensions this turns out to be impossible.
Therefore we keep the constraints as written above.

Due to these constraints the fields $\omega_\m{}^{ab}$, $f_\m{}^a$ and
$\phi_\m^i$ are no longer independent, but can be expressed in terms of
the other fields. In order to write down the explicit solutions of these
constraints, it is useful to extract the terms which have been underlined
in~(\ref{fcurv}). We define $\widehat R'$ as the curvatures without these
terms. Formally,
\begin{equation}
  \widehat R'_{\mu \nu }{}^I= \widehat R_{\mu \nu }{}^I+ 2h_{[\mu }^J
e_{\nu ]}^a
  f_{aJ}{}^I\,,
 \label{hatRprime}
\end{equation}
where $f_{aJ}{}^I$ are the structure constants in the $F^2(4)$ algebra
that define commutators of translations with other gauge transformations.
Then the solutions to the constraints are
 \bea \o^{ab}_\m
&=& 2 e^{\n[a} \partial_{[\m} e_{\n]}^{~b]} - e^{\n[a} e^{b]\s} e_{\m c}
\partial_\n e^{~c}_\s
 + 2 e_\m^{~~[a} b^{b]} - \ft12 \bar{\p}^{[b} \g^{a]} \p_\m - \ft14
\bar{\p}^b \g_\m \p^a \,,\nonumber\\
\f^i_\m &=& \ft13\rmi \g^a \widehat{R}^\prime _{\m a}{}^i(Q) - \ft1{24}\rmi
\g_\m \g^{ab} \widehat{R}^\prime _{ab}{}^i(Q)\,, \label{transfDepF} \\
f^a_\m &=& \ft16{\cal R}_\mu {}^a -\ft1{48}e_\mu {}^a {\cal R}\,,\qquad
{\cal R}_{\mu \nu }\equiv \widehat{R}_{\mu \rho }^{\prime~~ab}(M) e_a{}^\rho
e_{\nu b}\,,\qquad {\cal R}\equiv {\cal R}_\mu {}^\mu \,.\nonumber
 \eea
The constraints imply through Bianchi identities further relations
between the curvatures. The Bianchi identities for $R(P)$ imply
\begin{equation}
{\cal R}_{\mu \nu } = {\cal R}_{\nu \mu }\,, \qquad e_{[\m}{}^a \widehat
R_{\n\r ]}(D) = \widehat R_{[\m\n\r ]}{}^a(M)\,,\qquad
\widehat R_{\m\n}(D) = 0  \,. \label{Ricci}
\end{equation}

\subsection{Adding matter fields}

After imposing the constraints we are left with 21 bosonic and  24
fermionic degrees of freedom. The independent fields are those in the
left upper part of table~\ref{tbl:fieldsWeyls}. These have to be
completed with matter fields to obtain the full Weyl multiplet. We have
already seen that there are two possibilities for a $D=5$ Weyl multiplet
with each $32+32$ degrees of freedom.

These are obtained by adding either the left lower corner or right lower
corner of table~\ref{tbl:fieldsWeyls}. To obtain all the extra
transformations we imposed the superconformal algebra, but at the same
time allowing modifications of the algebra by field-dependent quantities.
The techniques are the same as already used in 4 and 6 dimensions
in~\cite{deWit:1980ug, Bergshoeff:1981is}, and were described in detail
in~\cite{Bergshoeff:1986mz}.

For the fields in the upper left corner, we now have to specify the extra
parts $M$ in~(\ref{transfogauge}). This will in fact only apply to
$Q$-supersymmetry. The other transformations are as in~(\ref{ftr}). The
extra terms we can read already from the linearized rules
in~(\ref{lintrDilW}) and~(\ref{lintrStW}). The full supersymmetry
transformations of these fields are
\begin{eqnarray}
 \d_Q e_\m{}^a   &=&  \ft 12\bar\e \g^a \psi_\m  \nn\, ,\\
\d_Q \psi_\m^i   &=& {\cal  D}_\m \e^i + \rmi \g\cdot T \g_\m \e^i  \nn\, ,\\
\d_Q V_\m{}^{ij} &=&  -\ft {3}{2}\rmi \bar\e^{(i} \phi_\m^{j)}+ \rmi
\bar\e^{(i} \g\cdot T \psi_\m^{j)} +4 \bar\e^{(i}\g_\m\chi^{j)}
    \nn\, ,\\
\d_Q b_\m       &=& \ft 12\rmi\bar\e\phi_\m -2 \bar\e\g_\mu \chi  \,,
\label{modifiedtransfgf}
\end{eqnarray}
where ${\cal D}_\mu \epsilon $ is given in~(\ref{covder2}). The fields
$T_{ab}$ and $\chi ^i$, and a further field $D$ that appears in their
transformation laws (see below) are independent fields in the Standard
Weyl multiplet, but not in the Dilaton Weyl multiplet. There, they are
given by expressions that are the non-linear extensions of~(\ref{linrelW}):
 \bea T_{ab}
&=& \ft18 \s^{-2}\bigg(\s \widehat{F}_{ab}
- \ft16 \ve_{abcde} \widehat{H}^{edc} + \ft14\rmi \bar{\p} \g_{ab} \p \bigg)
\, ,\label{TchiindilW} \nn \\
\chi^i &=& \ft18\rmi \s^{-1} \slashed{D} \p^i +\ft1{16}\rmi \s^{-2}
\slashed{D} \s \p^i - \ft1{32} \s^{-2} \g \cdot \widehat{F} \p^i+ \nn \\
& & + \ft14 \s^{-1} \g \cdot \underline{T} \p^i +
\ft1{32}\rmi \s^{-3}  \p_j\bar{\p}^i \p^j \,, \nn \\
D &=& \ft14 \s^{-1} \Box^c \s + \ft18 \s^{-2} (D_a \s) (D^a \s)  -\ft1{16}
\s^{-2}
\widehat{F}^2- \nn \\
& & - \ft18 \s^{-2} \bar{\p} \slashed{D} \p - \ft1{64} \s^{-4} \bar{\p}^i
\p^j \bar{\p}_i \p_j - 4 \rmi \s^{-1} \bar{\p} \underline{\chi}+ \nn \\
& & + \left(- \ft{26}3 \underline{T_{ab}} + 2 \s^{-1} \widehat{F}_{ab} +
\ft{1}{4} \rmi \s^{-2} \bar{\p} \g_{ab} \p \right) \underline{T^{ab}}\,,
\label{valueD}
 \eea
where the conformal d'alembertian is defined by
 \bea
\Box^c \s \equiv D^a D_a \s
&=& \left( \partial^a  - 2 b^a + \o_b^{~ba} \right) D_a \s - \ft12 \rmi
\bar{\p}_a D^a \p - 2  \s \bar{\p}_a \g^a \underline{\chi}+\nn \\
& & + \ft12 \bar{\p}_a \g^a \g \cdot \underline{T} \p + \ft12 \bar{\f}_a
\g^a \p + 2 f_a{}^a \s\,,
 \eea
and where the underlining indicates that these terms are dependent fields.
We have not substituted these terms in the expression for $D$ for reasons
of brevity.

The modification $M$ in~(\ref{transfogauge}) is the last term of the
transformations of $\psi _\mu ^i$, $V_\mu ^{ij}$ and $b_\mu $. The second
term in the transformation of $V_\mu ^{ij}$ on the other hand is due to
the fact that the structure constants have become structure functions,
and in particular there appears a new $T$-dependent $\SU(2)$
transformation in the anti-commutator of two supersymmetries. We will give
the full new algebra in section~\ref{ss:finalWeyls}.

The transformation rules for the matter fields\footnote{As we have already
seen, two of the extra fields in the Dilaton Weyl multiplet are actually
gauge fields, rather than matter fields. However, we use uniformly `matter
fields' for them in this context to indicate that they are not gauging a
symmetry of the superconformal algebra.} of the Weyl multiplets are as
follows. For the Standard Weyl multiplet we have ($Q$ and $S$
supersymmetry)
\begin{eqnarray}
\d T_{ab}     &=&  \ft {1}{2}\rmi \bar\e \g_{ab} \chi -\ft
{3}{32}\rmi \bar\e\widehat R_{ab}(Q)\,,    \nn\\
\d \chi^i     &=&  \ft 14 \e^i D -\ft{1}{64} \g\cdot \widehat R^{ij}(V) \e_j
                   + \ft18\rmi \g^{ab} \slashed{D} T_{ab} \e^i
                   - \ft18\rmi \g^a D^b T_{ab} \e^i-  \nn\\
               &&  -\ft 14 \g^{abcd} T_{ab}T_{cd} \e^i + \ft 16 T^2 \e^i
             +\ft 14 \g\cdot T \eta^i\, ,  \nn\\
\d D         &=&  \bar\e \slashed{D} \chi - \ft {5 }{3}\rmi \bar\e\g\cdot
T \chi - \rmi  \bar\eta\chi \,.
 \label{transfosmattStW}
\end{eqnarray}
There are no explicit gauge fields here, as should be the case for
`matter', i.e. non-gauge fields. These are all hidden in the covariant
derivatives and covariant curvatures. The covariant derivatives are for
any matter field given by the rule
\begin{equation}
  D_a \Phi=e_a^\mu \left( \partial _\mu -\delta _I(h_\mu ^I)\right)\Phi \,.
 \label{DaPhi}
\end{equation}
The last term represents thus a sum over all transformations except
general coordinate transformations, with parameters replaced by the
corresponding gauge fields. In practice, the Lorentz transformations and
$\SU(2)$ transformations follow directly from the index structure and lead
to additions similar to those in~(\ref{covder2}). For the Weyl
transformations there is a term $-w\,b_\mu \Phi $, where $w$ is the Weyl
weight of the field that can be found in table~\ref{tbl:fieldsWeyls}, and
then there remain the terms for $Q$ and $S$ supersymmetry. There are no
$K$ transformations for any matter field in this paper.

The covariant curvatures are given by the general rule (\ref{hatR}), e.g.
\begin{eqnarray}
  \widehat{R}_{\m\n}{}^i(Q) &=& R_{\m\n}{}^i(Q) + 2\rmi \gamma \cdot T
\g_{[\m} \p_{\n]}^i\,, \nonumber\\
  \widehat{R}_{\m\n}{}^{ij}(V)&=& R_{\m\n}{}^{ij}(V) - 8 \bar{\p}^{(i}_{[\m}
\g_{\n]} \chi^{j)} - \rmi \bar{\p}^{(i}_{[\m} \g\cdot T \psi_{\n]}^{j)}   \,,
 \label{hatRQV}
\end{eqnarray}
where $R_{\m\n}{}^i(Q)$ and $R_{\m\n}{}^{ij}(V) $ are those given
in~(\ref{fcurv}). Note that for $\widehat R(V)$ there are corrections
from modified structure functions as well as from $M$-dependent terms.
Having all the matter field dependence, we can obtain further
consequences of the curvature constraints. E.g.\ the Bianchi identity on
${\widehat R}(Q)$ gives:\footnote{We thank F. Coomans for drawing our attention to mistakes in versions v1 and v2 of this paper.}
\begin{eqnarray}
  \gamma \cdot\widehat{R}(S)&=&-\ft83T\cdot\widehat{R}(Q)\,,\nonumber\\
  \gamma ^\mu \widehat{R}_{\mu \nu }(S)&=&-\ft12\rmi D^\mu \widehat{R}_{\mu \nu }(Q)
  +\ft43\gamma _\nu T\cdot\widehat{R}(Q)+2\gamma _bT^{b\mu }\widehat{R}_{\mu \nu }(Q)
  \,,\nonumber\\
\widehat{R}_{\mu \nu }(S)&=&-\rmi \slashed{D}\widehat{R}_{\mu \nu }(Q)-\rmi\gamma _{[\mu }{\cal D}^\rho \widehat{R}_{\nu ]\rho }(Q)
-\ft83\gamma _{\mu \nu } T\cdot\widehat{R}(Q)\label{BIRQRS}\\
&&+12\gamma _bT^{\rho b}\gamma _{[\mu }\widehat{R}_{\nu ]\rho }(Q)-3\gamma \cdot T\widehat{R}_{\mu \nu }(Q)
-8T^\rho {}_{[\mu }\widehat{R}_{\nu ]\rho }(Q)\,.\nonumber
\end{eqnarray}

Given these transformation rules, we can calculate the transformations of
the dependent fields. Their transformation rules are now determined by
their definition due to the constraints. An equivalent way of expressing
this is that their transformation rules are modified w.r.t.~(\ref{ftr}),
due to the non-invariance of the constraints under these transformations.
We have chosen the constraints to be invariant under all bosonic
symmetries without modifications. Therefore, only the $Q$- and
$S$-supersymmetries of the dependent fields are modified to get invariant
constraints. The new transformation of the spin connection is
\begin{eqnarray}
\d\o_\m{}^{ab} &=& \ft12\rmi \bar\e \g^{ab}\phi_\m  - \ft12\rmi \bar
\eta \gamma ^{ab}\psi _\mu- \nonumber\\
 && - \rmi \bar\e \g^{[a} \g\cdot T \g^{b]} \psi_\m -\nn\\
 && -\ft 12 \bar\e\g^{[a} \widehat R_\m{}^{b]}(Q)
 - \ft 14 \bar\e\g_\m \widehat R^{ab}(Q)  - 4 e_\m{}^{[a} \bar\e \g^{b]} \chi
\,.
                \label{e:var omega}
\end{eqnarray}
The first line is the transformation as implied from the $F^2(4)$
algebra, see~(\ref{ftr}). The second line is due to the modification of
the anti-commutator of two supersymmetries by a $T$-dependent Lorentz
rotation. Finally, the last line contains the terms that go into the $M$
of~(\ref{transfogauge}). We give here for $\phi _\mu ^i$ just the latter
type of terms
 \bea
\d\phi_\mu^i    &=& \dots
%
% we leave out these soft algebra terms. I keep them after % if someone wants
% to recover them.
%-\ft 94 \rmi \bar\e\psi_\mu \chi^i + \ft 74 \rmi \bar\e\g_a\psi_\mu
%\g^a\chi^i + \ft 14 \rmi \bar\e^{(i}\g_{ab}\psi_\mu^{j)} \left( \g^{ab}\chi_j
%+ \ft 14 \widehat R^{cd}{}_j(Q) \right)        \nn\\
%&&
%
%
-\ft1{12}\rmi \left\{ \g^{ab}\g_\m - \ft 12 \g_\m\g^{ab}
                    \right\} \widehat R_{ab}{}^i{}_j(V) \e^j+  \nn\\
        &&  + \ft {1}{3}\left[ \slashed{D} \g\cdot T \g_\m - D_\mu
\g\cdot T
            + \g_\m\g^cD^a T_{ac} \right] \e^i+  \nn\\
        &&  + \rmi \left[  - \gamma _{\mu  abcd}T^{ab}T^{cd}
  +8 \gamma _\rho T^{\rho \sigma }T_{\mu \sigma } -2\gamma _\mu T^2 \right] \e^i+  \nn\\
        &&  + \ft13\rmi \left( 8 \g^b T_{\mu b} - \g_\m
\g\cdot T \right) \eta^i  \,.
          \label{e:var phi}
\eea We will not need the transformations for the field $f_\m{}^a$,
except the transformation of $f_a{}^a$ under $S$, since this term appears
in the conformal d'alembertian. We only give its $M$-dependent
$S$-transformation
\be
\d_S f_a{}^a = -5 \rmi \bar{\eta} \chi\,.
 \ee
We also used the transformation of the following curvatures:
 \bea
\d\widehat R_{ab}{}^i(Q) &=&  \ft 16 \left( \g_{ab}{}^{cd} -
\g^{cd}\g_{ab} - \ft 12 \g_{ab}\g^{cd}
                         \right) \widehat R_{cd}{}^i{}_j(V) \e^j
                     + \ft 14 \widehat R_{ab}{}^{cd}(M) \g_{cd} \e^i+  \nn\\
          && + 2\rmi \left(D_{[a} \g\cdot T \g_{b]} - \ft 13 D_{[a}
\g_{b]} \g\cdot T-\right.\nonumber\\
   &&\phantom{.}\hspace{1cm}\left.
    - \ft 13 \g_{[a} \slashed{D} \g\cdot T \g_{b]} - \ft 13
\g_{ab} D_c \g_d T^{cd}\right) \e^i \nonumber\\
&& -8T_{ab}\eta ^i+ \ft{16}3T_{[a}{}^c\gamma _{b]c}\eta ^i +\ft43 \gamma
_{abcd}
  T^{cd}\eta ^i\,, \label{deltaRQV}
\\
%%%%
\d\widehat R_{ab}{}^{ij}(V) &=& -\ft32\rmi \bar\e^{(i} \widehat R_{ab}{}^{j)}
(S)
-8 \bar\e^{(i}\g_{[a}D_{b]}\chi^{j)}
                        + \rmi \bar \epsilon ^i\gamma \cdot T
\widehat R_{ab}{}^{j)}(Q)+  \nn\\
                     && +8\rmi \bar\e^{(i}\g_{[a}\g\cdot T \g_{b]}\chi^{j)}
                        +\ft32\rmi\bar \eta ^{(i} \widehat R_{ab}{}^{j)}(Q)
                        + 8\rmi \bar \eta ^{(i} \gamma _{ab }\chi ^{j)}\,\nn .
\eea

The $Q$- and $S$-supersymmetry variations of the matter fields in the
Dilaton Weyl multiplet are
\begin{eqnarray}
\d A_\m
&=& -\ft12\rmi \s \bar{\e} \p_\m + \ft12
\bar{\e} \g_\m \p \, , \nonumber\\
 \d B_{\m\n}
&=& \ft12 \s^2 \bar{\e} \g_{[\m} \p_{\n]} + \ft12 \rmi \s \bar{\e} \g_{\m\n} \p
+ A_{[\m} \d(\e) A_{\n]}
\, , \nonumber\\
\d \p^i &=& - \ft14 \g \cdot \widehat{F} \e^i -\ft12\rmi \slashed{D} \s
\e^i + \s \g \cdot \underline{T} \e^i - \ft14 \rmi \s^{-1} \e_j
\bar{\p}^i \p^j+ \s
\eta^i \, , \nonumber\\
\d \s &=& \ft12 \rmi \bar{\e} \p\, .
 \label{transfmatterDilW}
\end{eqnarray}
The gauge fields $A_\m$ and $B_{\m\n}$ have the additional symmetries
 \bea
 \d A_\m &=& \partial_\m \L\,, \nonumber\\
\d B_{\m\n} &=&  2 \partial_{[\m} \L_{\n]} -\ft12 \L F_{\m\n}\,.
 \eea
Note that the dependence of the transformation rules for $A_\mu $ and
$B_{\mu \nu }$  on $\psi _\mu $ and $A_\mu $ signal new terms in the
algebra of supersymmetries and $U(1)$ transformations\footnote{The
$A_{[\mu} \psi_{\nu ]}$ term in $\delta B_{\mu \nu }$ is an extension
of~(\ref{transfogauge}) that occurs for antisymmetric tensor gauge fields.}.
This algebra will be written in section~\ref{ss:finalWeyls}. On the other
hand, the $F$-term in $\delta B_{\mu \nu }$ should be interpreted as an
$M$ term according to~(\ref{transfogauge}), and modifies the field
strength accordingly. This leads to the following field strengths of
these gauge fields \bea \widehat{F}_{\m\n} &=& 2\partial _{[\mu }A_{\nu
]} + \ft12\rmi \s \bar{\p}_{[\m} \p_{\n]}
 -  \bar{\p}_{[\m} \g_{\n]} \p\, , \nonumber\\
\widehat{H}_{\m\n\r} &=&3\partial _{[\mu }B_{\nu \rho ]} - \ft34 \s^2
\bar{\p}_{[\m} \g_\n \p_{\r]} - \ft32\rmi \s \bar{\p}_{[\m} \g_{\n\r]} \p
+ \ft32 A_{[\m} F_{\n\r]}\, . \label{hatFH}
 \eea
For the convenience of the reader we give their transformation rules:
 \bea \d \widehat{F}_{ab}
&=& - \ft12\rmi \s \bar{\e} \widehat{R}_{ab}(Q) - \bar{\e} \g_{[a} D_{b]} \p
 + \rmi \bar{\e} \g_{[a} \g \cdot \underline{T} \g_{b]} \p  +\rmi \bar{\eta} \g_{ab} \p \, , \nn \\
\d \widehat{H}_{abc}
&=& -\ft34 \s^2 \bar{\e} \g_{[a} \widehat{R}_{bc]}(Q) + \ft32\rmi \bar{\e}
\g_{[ab} D_{c]} \p + \ft32 \rmi D_{[a} \s \bar{\e} \g_{bc]}\p - \nn \\
&& - \ft32 \s \bar{\e} \g_{[a} \g \cdot \underline{T} \g_{bc]} \p -\ft32
\bar{\e} \g_{[a} \widehat{F}_{bc]} \p  - \ft32 \s \bar{\eta} \g_{abc} \p
\,.
 \eea
Finally, we give the Bianchi identities for these two curvatures
\bea D_{[a} \widehat{F}_{bc]}
&=& \ft12 \bar{\p} \g_{[a} \widehat{R}_{bc]}(Q)\,, \nn \\
D_{[a} \widehat{H}_{bcd]}
&=& \ft34 \widehat{F}_{[ab} \widehat{F}_{cd]}\,.
\eea

This finishes our discussion of the Standard and Dilaton Weyl multiplets.
The final results for these multiplets have been collected
in section~\ref{ss:finalWeyls}.
In the next section we will explain the connection between the
two multiplets.

\section{Results for the two Weyl multiplets}\label{ss:finalWeyls}

For the convenience of the reader we collect in this section the
essential results of the previous sections, and give the supersymmetry
algebra, which is modified by field-dependent terms. The transformation
under dilatation is for each field $\delta _D\Phi =w\Lambda _D\Phi $,
where the Weyl weight $w$ can be found in table~(\ref{tbl:fieldsWeyls}).
The Lorentz, and $\SU(2)$ transformations are evident from the index
structure, and our normalizations can be found in~(\ref{ftr}).

\subsection{The Standard Weyl multiplet}
The $Q$- and $S$-supersymmetry and $K$-transformation rules for the
independent fields of the Standard Weyl multiplet are
 \bea
\d e_\m{}^a   &=&  \ft 12\bar\e \g^a \psi_\m  \nn\, ,\\
\d \psi_\m^i   &=& {\cal  D}_\m \e^i + \rmi \g\cdot T \g_\m \e^i -\rmi
\g_\m
\eta^i  \nn\, ,\\
\d V_\m{}^{ij} &=&  -\ft32\rmi \bar\e^{(i} \phi_\m^{j)} +4
\bar\e^{(i}\g_\m\chi^{j)}
  + \rmi \bar\e^{(i} \g\cdot T \psi_\m^{j)} + \ft32\rmi
\bar\eta^{(i}\psi_\m^{j)} \nn\, ,\\
\d T_{ab}     &=&  \ft12\rmi \bar\e \g_{ab} \chi -\ft
{3}{32}\rmi \bar\e\widehat R_{ab}(Q)    \nn\, ,\\
\d \chi^i     &=&  \ft 14 \e^i D -\ft{1}{64} \g\cdot \widehat R^{ij}(V)
\e_j
                   + \ft18\rmi \g^{ab} \slashed{D} T_{ab} \e^i
                   - \ft18\rmi \g^a D^b T_{ab} \e^i - \nn\\
               &&  -\ft 14 \g^{abcd} T_{ab}T_{cd} \e^i + \ft 16 T^2 \e^i
             +\ft 14 \g\cdot T \eta^i  \nn\, ,\\
\d D         &=&  \bar\e \slashed{D} \chi - \ft {5}{3} \rmi
\bar\e\g\cdot T
\chi - \rmi  \bar\eta\chi \nn\, ,\\
\d b_\m       &=& \ft 12 \rmi\bar\e\phi_\m -2 \bar\e\g_\mu \chi +
\ft12\rmi \bar\eta\psi_\mu +2\Lambda _{K\mu } \,. \label{modifiedtransf}
 \eea
The covariant derivative ${\cal D}_\mu \epsilon $ is given
in~(\ref{covder2}). For other covariant derivatives, see the general
rule~(\ref{DaPhi}), with more explanation below that equation. The
covariant curvatures $\widehat{R}(Q)$ and $\widehat{R}(V)$ are given
explicitly in~(\ref{hatRQV}). The expressions for the dependent fields
are given in~(\ref{transfDepF}), where the prime indicates the omission
of the underlined terms in~(\ref{fcurv}).

\subsection{The Dilaton Weyl multiplet}
The Dilaton Weyl multiplet contains two extra gauge transformations: the
gauge transformations of $A_\mu $ with parameter $\Lambda $ and those of
$B_{\mu \nu }$ with parameter $\L _\mu $. The transformation of the fields
are given by:
 \bea
\d e_\m{}^a   &=&  \ft 12\bar\e \g^a \psi_\m  \nn\, ,\\
\d \psi_\m^i   &=& {\cal  D}_\m \e^i + \rmi \g\cdot \underline{T} \g_\m
\e^i - \rmi \g_\m
\eta^i  \nn\, ,\\
\d V_\m{}^{ij} &=&  -\ft32\rmi \bar\e^{(i} \phi_\m^{j)} +4
\bar\e^{(i}\g_\m \underline{\chi^{j)}}
  + \rmi \bar\e^{(i} \g\cdot \underline{T} \psi_\m^{j)} + \ft32\rmi
\bar\eta^{(i}\psi_\m^{j)} \nn\, ,\\
 \d A_\m
&=& -\ft12\rmi \s \bar{\e} \p_\m + \ft12
\bar{\e} \g_\m \p +\partial_\m \L \, , \nonumber\\
 \d B_{\m\n}
&=& \ft12 \s^2 \bar{\e} \g_{[\m} \p_{\n]} + \ft12 \rmi \s \bar{\e}
\g_{\m\n} \p + A_{[\m} \d(\e) A_{\n]}
+2 \partial_{[\m} \L_{\n]} -\ft12 \L F_{\m\n}\, , \nonumber\\
\d \p^i &=& - \ft14 \g \cdot \widehat{F} \e^i -\ft12\rmi \slashed{D} \s
\e^i + \s \g \cdot \underline{T} \e^i - \ft14 \rmi \s^{-1} \e_j \bar{\p}^i
\p^j + \s
\eta^i \, , \nonumber\\
\d \s &=& \ft12 \rmi \bar{\e} \p \, ,\nonumber\\
 \d b_\m       &=& \ft12 \rmi \bar\e\phi_\m -2 \bar\e\g_\mu \underline{\chi} +
\ft12\rmi \bar\eta\psi_\mu+2\Lambda _{K\mu } \,.
 \eea
The covariant curvature of $A_\m$ and $B_{\m\n}$ can be found
in~(\ref{hatFH}). The transformation of the dependent fields and the
curvatures have been given in the previous section. We have underlined
the fields $T_{ab}$ and $\chi^i$ to indicate that they are not
independent fields but merely short-hand notations. The explicit
expression for these fields in terms of fields of the Dilaton Weyl
multiplet are given in~(\ref{TchiindilW}).

\subsection{Modified superconformal algebra}
Finally, we present the `soft' algebra that these Weyl multiplets
realize. This is the algebra that all matter multiplets will have to
satisfy, apart from possibly additional transformations under which the
fields of the Weyl multiplets do not transform, and possibly field
equations if these matter multiplets are on-shell.

The full commutator of two supersymmetry transformations is
\begin{eqnarray}
 \left[\d_Q(\e_1),\d_Q(\e_2)\right] &=&  \d_{cgct}(\xi_3^\m)+
\d_M(\l^{ab}_3) + \d_S(\eta_3)  + \d_U(\l^{ij}_3) + \nonumber\\ &&
+\d_K(\L^a_{K3})+ \d_{U(1)}(\L_3) + \d_B(\L_{3\m}) \,. \label{algebraQQ}
\end{eqnarray}
The covariant general coordinate transformations have been defined
in~(\ref{defcgct}). The last two terms appear obviously only in the
Dilaton Weyl multiplet formulation. The parameters appearing
in~(\ref{algebraQQ}) are
 \bea
\xi^\m_3       &=& \ft 12 \bar\e_2 \g_\m \e_1 \,,\nn\\
\l^{ab}_3      &=& - \rmi \bar \epsilon _2\gamma ^{[a}\gamma \cdot T
\gamma ^{b]}\epsilon _1 \,, \nonumber\\
\lambda^{ij}_3 &=& \rmi \bar\e^{(i}_2 \g\cdot T \e^{j)}_1\,, \nn\\
\eta^i_3       &=& - \ft {9}{4}\rmi\, \bar \e_2 \e_1 \chi^i
                   +\ft {7}{4}\rmi\, \bar \e_2 \g_c \e_1 \g^c \chi ^i+ \nn\\
               &&  + \ft1{4}\rmi\,  \bar \e_2^{(i} \g_{cd} \e_1^{j)}
\left( \g^{cd} \chi_j
                   + \ft 14\, \widehat R^{cd}{}_j(Q) \right) \, ,\nonumber\\
\Lambda^a_{K3} &=& -\ft 12 \bar\e_2\g^a\e_1 D + \ft{1}{96}
\bar\e^i_2\g^{abc}\e^j_1 \widehat R_{bcij}(V) + \nn\\
               && + \ft1{12}\rmi\bar\e_2\left(-5\g^{abcd} D_b T_{cd} +
9 D_b T^{ba} \right)\e_1+  \nn\\
               && + \bar\e_2\left(  \g^{abcde}T_{bc}T_{de}
                  - 4 \g^c T_{cd} T^{ad} +  \ft 23  \g^a T^2
                  \right)\e_1 \,,\nonumber\\
\L_3&=& -\ft12\rmi \s \bar{\e}_2 \e_1 \, ,\nonumber\\
\L_{3\m} &=& -\ft12 \s^2 \xi_{3\m} - \ft12 A_\m \L_3\,.
 \eea
For the $Q,S$ commutators we find the following algebra:
 \bea
 \left[\d_S(\eta),
\d_Q(\e)\right] &=& \d_D( \ft12\rmi \bar\e\eta ) + \d_M( \ft12\rmi \bar\e
\g^{ab} \eta) +      \d_U(  -\ft32\rmi \bar\e^{(i} \eta^{j)} )  +
\delta_K(\L_{3K}^a ) \,,\nn\\
\left[ \d_S(\eta_1), \d_S(\eta_2) \right] &=& \d_K( \ft 12 \bar\eta_2 \g^a
\eta_1 ) \,. \eea
 with
 \be
 \L_{3K}^a= \ft16 \bar{\e} \left(\g \cdot T \g_a - \ft12 \g_a \g \cdot T
\right) \eta \,. \label{algebraQSS}
\end{equation}
The commutator of $Q$ and $U(1)$ transformations is given by
\begin{equation}
 \left[\d(\e), \d(\L) \right] =\d_B\left(-\ft12 \L\d(\e)
A_\m \right) .
 \label{commQU1}
\end{equation}

This concludes our description of the Standard and Dilaton Weyl
multiplets.

\section{Connection between the Weyl multiplets}\label{ss:connection}

In the previous section we have shown that the Standard and Dilaton Weyl
multiplets can be related to each other by expressing the fields of the
Standard Weyl multiplet in terms of those of the Dilaton Weyl multiplet
(see~(\ref{valueD})). It is known that in 6 dimensions the coupling of an
on-shell self\-dual tensor multiplet to the $D=6$ Standard Weyl multiplet
leads to a $D=6$ Dilaton Weyl multiplet~\cite{Bergshoeff:1986mz}. Since
in 5 dimensions a tensor multiplet
 is dual to a vector multiplet, it is natural to consider
the coupling of a vector multiplet to the Standard Weyl multiplet. Since
the Standard Weyl multiplet has no dilaton we must consider the improved
vector multiplet. We will take the vector multiplet off-shell to simplify
the higher-order fermion terms.

\subsection{The improved vector multiplet}

We will first consider the improved vector multiplet in a flat background,
i.e.\ no coupling to conformal supergravity. Our starting point
is the lagrangian corresponding to an off-shell vector multiplet:
\be
{\cal L} = -{1 \over 4} F_{\m\n}F^{\m\n} - {1 \over 2} \bar{\p}
\slashed{\partial} \p - {1 \over 2} (\partial \s)^2 + Y^{ij} Y_{ij} \,.
\ee
The action corresponding to this lagrangian
 is invariant under the off-shell supersymmetries
\bea
\d A_\m
&=& \ft12
\bar{\e} \g_\m \p \, , \nonumber\\
\d Y^{ij}
&=& -\ft12 \bar{\e}^{(i} \slashed{\partial} \p^{j)}\,, \nonumber \\
\d \p^i
&=& - \ft14 \g \cdot F \e^i -\ft12\rmi \slashed{\partial} \s \e^i -  Y^{ij}
\e_j \, , \nonumber\\
\d \s &=& \ft12 \rmi \bar{\e} \p\,. \label{offshellvec}
\eea
The action has the wrong Weyl weight to be scale invariant. We therefore
improve it by multiplying all terms with the dilaton.
This requires additional cubic
terms in the action to keep it invariant under supersymmetry. We thus
obtain the lagrangian for the improved vector multiplet:
\bea
\label{fstl}
{\cal L}
&=& -\ft14 \s F_{\m\n} F^{\m\n} - \ft12 \s \bar{\p} \slashed{\partial} \p -
\ft12 \s (\partial \s)^2 + \s Y^{ij} Y_{ij}- \nn \\
& & -\ft18 \rmi \bar{\p} \g \cdot F \p -\ft1{24} \ve_{\m\n\l\r\s} A^\m
F^{\n\l} F^{\r\s} -\ft12 \rmi \bar{\p}^i \p^j Y_{ij} \,.
\eea
If we define the following
\bea
S^{ij} &=& 2 \s Y^{ij} - \ft12 \rmi \bar{\p}^i \p^j \,, \nn \\
\Gamma^i &=& \rmi \s \slashed{\partial} \p^i + \ft12 \rmi \slashed{\partial} \s \p^i - \ft14 \g \cdot F \p^i + Y^{ij} \p_j \,, \nn \\
C &=& -\ft14 F_{\m\n} F^{\m\n} - \ft12 \bar{\p} \slashed{\partial} \p + \s
\Box \s + \ft12 (\partial \s)^2 + Y^{ij} Y_{ij} \,, \nn \\
H_a &=& -\ft1{8} \ve_{abcde} F^{bc} F^{de} - \partial^b \left(- \s
F_{ba} - \ft14 \rmi \bar{\p} \g_{ba} \p \right) \,, \nn \\
G_{abc} &=& \partial_{[a} F_{bc]} \,, \label{flateom}
\eea
then the equations of motion and the Bianchi identity corresponding to this
lagrangian are given by
\be
0 = S^{ij} = \Gamma^i = C = H_a = G_{abc}\,.
\ee

\subsection{Coupling to the Standard Weyl multiplet}

Next, we consider the coupling of the improved vector multiplet to the
Standard Weyl multiplet. The transformation rules for the fields of the
off-shell vector multiplet can be found by imposing the superconformal
algebra~(\ref{algebraQQ}). We thus find the following $Q$- and
$S$-transformation rules:
\begin{eqnarray}
\d A_\m
&=& -\ft12\rmi \s \bar{\e} \p_\m + \ft12
\bar{\e} \g_\m \p \, , \nonumber\\
\d Y^{ij}
&=& -\ft12 \bar{\e}^{(i} \slashed{D} \p^{j)} + \ft12 \rmi \bar{\e}^{(i} \g
\cdot T \p^{j)} - 4 \rmi \s \bar{\e}^{(i} \chi^{j)} + \ft12 \rmi
\bar{\eta}^{(i} \p^{j)} \,, \nonumber \\
\d \p^i
&=& - \ft14 \g \cdot \widehat{F} \e^i -\ft12\rmi \slashed{D} \s \e^i + \s
\g \cdot T \e^i - Y^{ij} \e_j + \s \eta^i \, , \nonumber\\
\d \s &=& \ft12 \rmi \bar{\e} \p \,, \label{vectconform}
\end{eqnarray}
where the covariant curvature is
\begin{equation}
\widehat{F}_{\m\n} = 2\partial _{[\mu }A_{\nu ]} + \ft12\rmi \s \bar{\p}_{[\m}
\p_{\n]}
 -  \bar{\p}_{[\m} \g_{\n]} \p\,.
\end{equation}
The supercovariant extension of the Bianchi identity reads
\be
0 = G_{abc} = D_{[a} \widehat{F}_{bc]} - \ft12 \bar{\p} \g_{[a}
\widehat{R}_{bc]}(Q) \,. \label{covBI}
\ee

The first term in the transformation of $A_\mu $, reflected also in the
curvature, signals a modification of the supersymmetry algebra, as can be
seen by comparing with the general rule (\ref{transfogauge}):
\begin{equation}
 [ \d(\e_1) , \d(\e_2)] = \ldots + \d_{U(1)}\left(\L_3= -\ft12\rmi \s
\bar\e_2 \e_1 \right),
 \label{QQVM}
\end{equation}
where the dots indicate all the terms present for the fields of the Standard
 Weyl multiplet and where the last term is the gauge transformation of $A_\mu
 $. This $U(1)$ is not part of the superconformal algebra and has no effect on
 the fields of the Standard Weyl multiplet. This is similar to the central
 charge induced in vector multiplets in 4 dimensions.

Our next goal is to find the equations of motion for the improved vector
multiplet. These equations of motion should be an extension of the
flat spacetime results given in~(\ref{flateom}). One way to proceed is to
first find the curved background extension of the flat spacetime action
defined by~(\ref{fstl}) and next derive the equations of motion from
this action. However, for our present purposes, it is
sufficient to find the equations of motion only.

We want to identify the spinor $\psi^i$ of the vector multiplet with the
spinor $\psi^i$ of the Dilaton Weyl multiplet. This is why we have given
these two spinors the same name in the first place (see the footnote in
subsection~\ref{ss:lindilW}). Comparing the $\SU(2)$ triplet term in the
supersymmetry transformations of the two spinors,
see~(\ref{transfmatterDilW}) and~(\ref{vectconform}), we deduce that the
constraint $S^{ij}$ does not get any corrections and we must have
 \be
S^{ij} = 2 \s Y^{ij} - \ft12 \rmi \bar{\p}^i \p^j\,. \label{Sij}
 \ee
There are now two ways to proceed. One way is to make the transition to
an on-shell vector multiplet by using~(\ref{Sij}) to eliminate the
auxiliary field $Y^{ij}$ from the transformation
rules~(\ref{vectconform}). The commutator of two supersymmetry
transformations would then only close modulo the equations of motion.

A more elegant way is to note that the equations of motion must transform into each other. By varying~(\ref{Sij}) under~(\ref{vectconform}) we find
\bea
\d S^{ij} = \rmi \bar{\e}^{(i} \Gamma^{j)} \, ,
\eea
where the supercovariant extension of $\Gamma^i$ is now given by
\bea
\Gamma^i &=& \rmi \s \slashed{D} \p^i + \ft12 \rmi \slashed{D} \s \p^i - \ft14 \g \cdot \widehat{F} \p^i + Y^{ij} \p_j+ \nn \\
&& + 2 \s \g \cdot T \p^i - 8 \s^2 \chi^i \,. \label{gammacov}
 \eea
Varying this expression under~(\ref{vectconform}) and using~(\ref{covBI})
leads to the other equations of motion. We find:
 \be \d \Gamma^i =
-\ft12 \rmi \slashed{D} S^{ij} \e_j - \ft12 \rmi \g \cdot H \e^i + \ft12 C
\e^i - \g \cdot T S^{ij} \e_j \,,
 \ee
where the supercovariant generalizations of (\ref{flateom}) are given by
\bea C &=& -\ft14 \widehat{F}_{ab} \widehat{F}^{ab} - \ft12 \bar{\p}
\slashed{D}
\p + \s \Box^c \s + \ft12 D^a \s D_a \s + Y^{ij} Y_{ij}+ \nn \\
&& + \rmi \bar{\p} \g \cdot T \p - 16 \rmi \s \bar{\p} \chi -\ft{104}{3}
\s^2 T_{ab} T^{ab} + 8 \s \widehat{F}_{ab} T^{ab} -4 \s^2 D \,, \nn \\
H_a &=& -\ft1{8} \ve_{abcde} \widehat{F}^{bc} \widehat{F}^{de} - D^b
\left(8 \s^2 T_{ba} - \s \widehat{F}_{ba} - \ft14 \rmi \bar{\p} \g_{ba}
\p \right) \,. \label{curveom}
 \eea
The supercovariant equations of motion and Bianchi identity are then
given by \be 0 = S^{ij} = \Gamma^i = C = H_a = G_{abc}\,. \ee

\subsection{Solving the equations of motion}

In 6 dimensions, the equations of motion for an on-shell tensor multiplet
coupled to the Standard Weyl multiplet can be used to eliminate the matter
fields of the latter in terms of the matter fields of the Dilaton Weyl
multiplet. Precisely the same happens here. First of all the equations of
motion
 for $Y^{ij}$ can be used to eliminate this auxiliary field. Next,
the equations of motion for $\p^i$ and $\s$ can be used to solve for
the fields $\chi^i$ and $D$, respectively. The expressions for these fields
exactly coincide with the ones we found in~(\ref{valueD}).

The solution for the matter field $T_{ab}$ in terms of the fields of the
Dilaton Weyl multiplet is more subtle. It requires that we first
reinterpret the equation of motion for the vector field as the Bianchi
identity for a two-form antisymmetric tensor gauge field $B_{\mu\nu}$. To
be precise, we rewrite $H_a=0$ from~(\ref{curveom}) as a Bianchi identity
 \be \label{bid} D_{[a}
\widehat{H}_{bcd]} = \ft34 \widehat{F}_{[ab} \widehat{F}_{cd]}\,,
\label{BIH}
 \ee
where the three-form curvature $\widehat{H}_{abc}$ is defined by
 \be
\label{dr} -\ft16 \ve_{abcde} \widehat{H}^{edc} = 8 \s^2 T_{ab} - \s
\widehat{F}_{ab}
 - \ft14 \rmi \bar{\p} \g_{ab} \p \,.
 \ee
Note that the latter equation is just a rewriting of the
relation~(\ref{valueD}) we found in section~\ref{ss:gaugingSC}.

The Bianchi identity~(\ref{bid}) can be solved in terms of an
antisymmetric two-form gauge field $B_{\mu\nu}$. The superconformal
algebra~(\ref{QQVM}) imposes that such a field transforms under
supersymmetry as follows:
 \be \label{transfB}
  \d_Q B_{\m\n} = \ft12 \s^2
\bar{\e} \g_{[\m} \p_{\n]} + \ft12 \rmi \s \bar{\e} \g_{\m\n} \p +
A_{[\m} \d(\e) A_{\n]}\,.
 \ee
In addition one finds that the field $B_{\m\n}$ transforms under a
$\UU(1)$ and a vector gauge transformation as follows
 \be \d B_{\m\n} = 2 \partial_{[\m} \L_{\n]} -\ft12 \L F_{\m\n}
\,. \label{dBmn}
 \ee
Furthermore, the commutator of two $Q$-transformations picks up a vector
gauge transformation $\delta_B$ for the field $B_{\m\n}$:
 \bea
&&[ \d(\e_1) , \d(\e_2)] = \ldots + \d_{U(1)}\left(\L_3\right) + \d_B
\left(\L_{3\m}\right)\,,\label{QQVTM} \nn \\
&&\L_3 = -\ft12\rmi \s \bar\e_2 \e_1 \,, \qquad \L_{3\m} = -\ft14 \s^2
\bar{\e}_2 \g_\m \e_1 - \ft12 A_\m \L_3 \,. \eea {}From the
transformation rules~(\ref{dBmn}) for $B_{\m\n}$ it follows that the
supercovariant field strength $\widehat{H}_{\m\n\r}$ is given by
 \be
\widehat{H}_{\m\n\r} = 3\partial _{[\mu }B_{\nu \rho ]} - \ft34 \s^2
\bar{\p}_{[\m} \g_\n \p_{\r]} - \ft32\rmi \s \bar{\p}_{[\m} \g_{\n\r]} \p
+ \ft32 A_{[\m} F_{\n\r]}\, .
 \ee
This field strength indeed satisfies the Bianchi identity~(\ref{BIH}).

We conclude that the connection between the Standard and Dilaton Weyl
multiplets can be obtained by first coupling an improved vector multiplet
to the Standard Weyl multiplet and, next, solving the equations
of motion. To solve for the equation of motion for the vector field
in terms of the matter field $T_{ab}$
one must first reinterpret this equation of motion as
the Bianchi identity for an antisymmetric two-form gauge field.

\section{Conclusions}\label{ss:conclusions}

In this work we have taken the first step in the superconformal tensor
calculus by constructing the Weyl multiplets for $N=2$ conformal supergravity
 theory in 5 dimensions.

First, we have applied the standard current multiplet procedure to the
case of the $D=5$ vector multiplet. An unconventional feature is that the
corresponding energy-momentum tensor is neither traceless nor improvable
to a traceless current. However, since one version
 of the Weyl multiplet contains a dilaton we could construct this linearized
$32+32$ Dilaton Weyl multiplet from the current multiplet. We also pointed
out that there exists a second (`Standard') Weyl multiplet without a
dilaton, by comparing with similar Weyl multiplets in $D=4$ and $D=6$.

Next, we explained how the non-linear multiplets could be obtained by
gauging the superconformal algebra $F^2(4)$. Finally, we discussed the
relation between the two Weyl multiplets. We showed that the coupling of
the Standard Weyl multiplet to an improved vector multiplet leads to the
non-linear relation between the Standard and Dilaton Weyl multiplets.

The fact that there exist two different versions of conformal
supergravity has been encountered before in 6
dimensions~\cite{Bergshoeff:1986mz}. Table~\ref{tbl:compWeyl456} suggests
that the same feature might also occur in 4 dimensions.
\begin{table}[htb]
\begin{center}
\begin{tabular}{||c|c|c|c||}
\hline
\rule[-1mm]{0mm}{6mm}
Dimension $D$      &  \# d.o.f.    & Standard Weyl    & Dilaton Weyl \\
\hline
\rule[-1mm]{0mm}{6mm}
6&10&$T_{abc}^+$&$B_{\mu\nu}$\\
\rule[-1mm]{0mm}{6mm}
5&10&$T_{ab}$&$A_\mu\, , B_{\mu\nu}$\\
\rule[-1mm]{0mm}{6mm}
4&6&$T_{ab}$&$A_\mu\, , B_\mu$\\
\hline
\end{tabular}
\caption{\it The two different formulations of the Weyl multiplet in
$D=4,5,6$.}\label{tbl:compWeyl456}
\end{center}
\end{table}
It seems plausible that in this case the coupling of a vector multiplet to
the Standard Weyl multiplet will give a Dilaton Weyl multiplet containing
two vectors. It would be interesting to see whether the Dilaton Weyl
multiplet in 4 dimensions indeed exists, and how it can be used in matter
couplings.

It is known, from the AdS/CFT correspondence, that there is a relation
between $(1,1)$, $D=6$ gauged supergravity with 16 supercharges and the
$N=2, D=5$ conformal supergravity with 8 supercharges. In fact, the
precise relation between the Standard Weyl multiplet and the $(1,1), D=6$,
or $F(4)$-gauged, supergravity~\cite{Romans:1986tw} has been given
in~\cite{Nishimura:2000wj}. The $F(4)$-gauged supergravity contains a {\it
massive} antisymmetric two-form tensor which, according
to~\cite{Nishimura:2000wj}, corresponds to the $T_{ab}$ matter field of
the Standard Weyl multiplet. It would be interesting to see whether the
work of~\cite{Nishimura:2000wj} can be extended such that the
$F(4)$-gauged supergravity theory also gives rise to the Dilaton Weyl
multiplet. One possibility is that, in order to achieve this, one should
first replace the massive two-form of gauged supergravity by a {\it
massless} one-form and two-form gauge field.

Finally, the results of this work will be our starting point for the
construction of general supergravity/matter couplings in 5 dimensions. We
hope to report about this in the nearby future.

\bigskip

\centerline{NOTE ADDED}

A few days after we sent this paper to the bulletin board an interesting
paper appeared on conformal supergravity in five
dimensions~\cite{Fujita:2001kv} that has some overlap with our work. The
authors of~\cite{Fujita:2001kv} also discuss the two versions of the Weyl
multiplet. In addition they discuss the superconformal tensor calculus in
five dimensions.

\section*{Acknowledgements}
\noindent We like to thank S.~Vandoren and F.~Roose for useful
discussions and especially M.~de Roo for discussions and his
computer-aided help with some of the difficult $\gamma$-calculations.
This work is supported by the European Commission RTN programme
HPRN-CT-2000-00131, in which E.B., R.H. and T.de W. are associated to the
university of Utrecht. R.H.\ was in part supported by a Marie Curie
fellowship of the European Community programme ``Improving Human Research
Potential and the Socio-Economic Knowledge Base'', contract number
HPMT-CT-2000-00165. We were also supported by the programme
\emph{Supergravity, Superstrings and M-Theory} of the Centre \'Emile
Borel of the Institut Henri Poincar{\'e} Paris (UMS 839-CNRS/UPMC), and
thank this center and the Spinoza institute in Utrecht for hospitality.
A.V.P. thanks the Caltech-USC Center for hospitality while the paper was
finalized.

\appendix

\section{Notations and Conventions}\label{app:conventions}
The metric is $(-++++)$, and we use the following indices (spinor indices are
always omitted)
\begin{eqnarray}
 \mu  & 0,\dots ,4 & \mbox{local spacetime} \,,    \nonumber\\
 a & 0,\dots ,4 & \mbox{tangent spacetime}\,,     \nonumber\\
  i  & 1,2 & \SU(2) \,.
\label{indices}
\end{eqnarray}

The generators $U_{ij}$ of the $R$-symmetry group $\SU(2)$ are defined to
be anti-hermitian and symmetric, i.e.
\be (U_i{}^j)^*=-U_j{}^i\,, \qq
U_{ij} = U_{ji}\,.
\ee A symmetric traceless $U_i{}^j$
corresponds to a symmetric $U^{ij}$ since we lower or raise $\SU(2)$ indices
 using the $\ve$-symbol, in NW--SE convention:
\begin{equation}
  X^i=\varepsilon ^{ij}X_j\,,\qquad X_i=X^j\varepsilon _{ji}\,,\qquad
\varepsilon _{12}=-\varepsilon _{21}=
\varepsilon ^{12}=1\,.
\label{NWSEconv}
\end{equation}
The actual value of $\varepsilon $ is here given as an example. It is in
fact arbitrary as long as it is antisymmetric, $\varepsilon
^{ij}=(\varepsilon _{ij})^*$ and $\varepsilon _{jk}\varepsilon
^{ik}=\delta_j{}^i$.

The charge conjugation matrix ${\cal C}$ and  ${\cal C}\gamma _a$ are
antisymmetric. The matrix  ${\cal C}$ is unitary and $\gamma _a$ is
hermitian apart from the timelike one, which is anti-hermitian. The bar
is the Majorana bar:
\begin{equation}
  \bar \lambda ^i =(\lambda ^i) ^T {\cal C}\,.
\label{barlambda}
\end{equation}
We define the charge conjugation operation on spinors as
\begin{equation}
(  \lambda ^i)^C\equiv \alpha^{-1}B^{-1}\varepsilon ^{ij}(\lambda
^j)^*\,, \qquad \bar \lambda ^{iC}\equiv \overline {(  \lambda ^i)^C}=
\alpha ^{-1}\left( \bar \lambda{}^k\right) ^*B\varepsilon ^{ki}\,,
 \label{defCspinors}
\end{equation}
where $B={\cal C}\gamma _0$, and $\alpha = \pm 1$ when one
uses the convention that complex conjugation does not interchange the order
of spinors, or $\alpha = \pm\rmi$ when it does.
Symplectic Majorana spinors satisfy
 $\lambda =\lambda
^C$. Charge conjugation acts on gamma matrices as $(\gamma _a)^C=-\gamma
_a$, does not change the order of matrices, and works on matrices in
$\SU(2)$ space as $M^C=\sigma _2 M^*\sigma _2$. Complex conjugation can
then be replaced by charge conjugation, if for every bispinor one inserts
a factor $-1$. Then, e.g.\ the expressions
\begin{equation}
  \bar \lambda ^i \gamma _\mu \lambda^j \,,\qquad \rmi \bar{\l}^i \chi _i
\label{realexpr}
\end{equation}
are real for symplectic Majorana spinors. For more details, see
e.g.~\cite{VanProeyen:1999ni}.

When the $\SU(2)$ indices on spinors are omitted, northwest-southeast
 contraction is
understood, e.g.
\be
{\bar\l} \g^{(n)} \psi = {\bar\l}^i \g^{(n)} \psi_i\,,
\ee
where we have used the following notation
\be
\g^{(n)} = \g^{a_1\cdots a_n} =\g^{[a_1}\g^{a_2}\cdots \g^{a_n]}\,.
\ee

The anti-symmetrizations are always with unit strength. Changing the
order of spinors in a bilinear leads to the following signs
\be
{\bar \psi}^{(1)} \g_{(n)} \chi ^{(2)} =
t_n
\ {\bar
\chi }^{(2)} \g_{(n)} \psi^{(1)}\ ,\qquad
\left\{ \begin{array}{c}
  t_n=-1\mbox{ for }n=2,3 \\
  t_n=+1\mbox{ for }n=0,1
\end{array}\right.
\ee
 where the labels $(1)$ and $(2)$ denote any $\SU(2)$ representation.

We frequently use the following Fierz rearrangement formulae
\be
\psi_j \bar{\l}^i = - \ft 14 \bar{\l}^i \psi_j -\ft14 \bar{\l}^i \g^a \psi_j
\g_a + \ft18 \bar{\l}^i \g^{ab} \psi_j \g_{ab}\,,\qquad \bar{\psi}^{[i}
\l^{j]} = - \ft 12 \bar{\psi} \l \ve^{ij}\, .
\ee

When one multiplies three spinor doublets, one should be able to write
the result in terms of $(8\times 7\times 6)/3!=56$ independent structures.
{}From analyzing the representations, one can obtain that these are in
the $(4,2)+(4,4)+(16,2)$ representations of $\overline{\SO(5)}\times \SU(2)$.
They are
\begin{eqnarray}
   && \xi _j\bar \xi ^j\xi ^i= \gamma ^a\xi _j\bar \xi ^j\gamma _a\xi
^i=\ft18\gamma ^{ab} \xi ^i \bar \xi \gamma
  _{ab}\xi\,, \nonumber\\
   && \xi ^{(k} \bar \xi ^{i}\xi ^{j)}\,,\nonumber\\
   && \xi _j\bar \xi ^j\gamma _a\xi ^i\,.
 \label{cubicfermions}
\end{eqnarray}

The Levi--Civit\`{a} tensor is real and satisfies
\begin{equation}
  \varepsilon _{a_1 \ldots a_n b_1 \ldots b_p} \varepsilon^{a_1 \ldots a_n
c_1 \ldots c_p} =- n! p! \d_{[b_1}^{[c_1} \ldots \d_{b_p]}^{c_p]}\,,\qquad
  \varepsilon ^{\mu \nu \rho \sigma\tau }=ee^\mu _ae^\nu _b\cdots
e^\tau _e\varepsilon
^{abcde}\,.
 \label{LeviCivita}
\end{equation}
We introduce the dual of a tensor as
\begin{equation}
{\tilde   A}{}^{a_1\ldots a_{5-n}}=\ft1{n!}\rmi \varepsilon _{a_1\ldots
a_{5-n}b_1\ldots b_n} A^{b_n\ldots b_1}\,,
 \label{dualin5}
\end{equation}
with the properties
\begin{equation}
  \tilde {\tilde A}=A\,, \qquad \frac{1}{n!}A^{a_1\ldots a_n}B_{a_1\ldots
  a_n}=\frac{1}{n!}A\cdot B=\frac{1}{(n-5)!}\tilde A\cdot \tilde B\,,
 \label{tildetilde}
\end{equation}
where we have introduced the notation $A\cdot B$ that we use throughout
the paper.

The product of all gamma matrices is proportional to the unit matrix in
odd dimensions. We use
\begin{equation}  \gamma ^{abcde}=\rmi\varepsilon^{abcde}\,.
\label{gammaepsilon}
\end{equation}
This implies that the dual of a $(5-n)$-antisymmetric gamma matrix is the
$n$-antisymmetric gamma matrix given by \be \label{gamma_dual}
\g_{a_1\ldots a_{n}} = \ft1{(5-n)!} \rmi \ve_{a_1\ldots a_{n} b_1 \ldots
b_{5-n}} \g^{b_{5-n} \ldots b_1}\,. \ee

For convenience we will give a rule for calculating gamma-contractions
like \be \label{contractions} \g^{(m)} \g_{(n)} \g_{(m)} = c_{n,m}
\g_{(n)}\, , \ee where the constants $c_{n,m}$ are given for the most
frequently used cases in table~\ref{tbl:contractions}.

\begin{table}[ht]
\begin{center}
\begin{tabular}{|l|rr|} \hline\hline
$d=5$  &  $m=1$  &  $m=2$ \\ \hline
$n=0$  &  $5$    &  $-20$ \\
$n=1$  &  $-3$   &  $-4$  \\
$n=2$  &  $1$    &  $4$   \\
$n=3$  &  $1$    &  $4$   \\ \hline
\end{tabular}
\caption{The coefficients $c_{n,m}$ in~(\ref{contractions}).}
\label{tbl:contractions}
\end{center}
\end{table}

\section{The $D=5$ superconformal algebra $F^2(4)$}\label{app:algF4}
There exist many varieties of superconformal algebras, when one allows
for central charges~\cite{vanHolten:1982mx,D'Auria:2000ec}. However, so
far a suitable superconformal Weyl multiplet has only been constructed
from those superconformal algebras\footnote{One notable case is the 10
dimensional Weyl multiplet~\cite{Bergshoeff:1983az}, that is not based on
a known algebra.} that appear in the Nahm's
classification~\cite{Nahm:1978tg}. In that classification appears one
exceptional algebra, which is $F(4)$. The particular real form that we
need here is denoted by $F^2(4)$, see tables 5 and 6
in~\cite{VanProeyen:1999ni}.

The commutation relations defining the $F^2(4)$ algebra are given by
\begin{equation}
  \begin{array}{rclrcl}
\left[P_a   , M_{bc}\right]  & = & \eta_{a[b}P_{c]}\,,\qquad  &  \left[K_a   , M_{bc} \right] & = & \eta_{a[b} K_{c]} \, ,  \\
\left[D     , P_a \right]    & = & P_a\, , \qquad &
\left[D     , K_a \right]    & = & -K_a\, ,  \\
\left[M_{ab}, M^{cd}\right]  & = & -2 \d_{[a}{}^{[c} M_{b]}{}^{d]}\,
,\qquad
  & \left[P_a   , K_b  \right]   & = & 2(\eta_{ab} D + 2 M_{ab})\, , \\
  &  &  &   &  &  \\
\left[M_{ab}, Q_{i\a} \right] & = & -\ft 14 (\g_{ab} Q_i)_\a \, ,\qquad  &
\left[M_{ab}, S_{i\a} \right] & = & -\ft 14 (\g_{ab} S_i)_\a \, , \\
\left[D     , Q_{i\a} \right] & = &  \ft 12 Q_{i\a}\, , &
\left[D     , S_{i\a} \right] & = & -\ft 12 S_{i\a}\, ,   \\
\left[K_a   , Q_{i\a} \right] & = &  \rmi(\g_a S_i)_\a \, , \qquad &
\left[P_a   , S_{i\a} \right] & = & -\rmi(\g_a Q_i)_\a \, ,   \\
  &  &  &  &  &  \\
\left\{ Q_{i\a}, Q_{j\b} \right\} & = & -\ft 12 \ve_{ij} (\g^a)_{\a\b} P_a
\, ,\qquad &\left\{ S_{i\a}, S_{j\b} \right\} & = & -\ft 12 \ve_{ij}
(\g^a)_{\a\b} K_a
\, ,   \\
\left\{ Q_{i\a}, S_{j\b} \right\} & = &\multicolumn{4}{l}{ -\ft 12 \rmi
\left(\ve_{ij} C_{\a\b}
 D +
      \ve_{ij} (\g^{ab})_{\a\b} M_{ab}  + 3 C_{\a\b} U_{ij} \right)
\, ,}  \\
  &  &  &  &  &   \\
\left[Q_{i\a}, U_{kl} \right] & = & \ve_{i(k} Q_{l)\a} \, , \qquad &
\left[S_{i\a}, U_{kl} \right] & = & \ve_{i(k} S_{l)\a} \, , \\
\left[U_{ij}, U^{kl}  \right] & = & 2 \d_{(i}{}^{(k} U_{j)}{}^{l)} \, .
 &  &  &
\end{array}
\label{suf2(4)}
\end{equation}
The first six commutation relations define the bosonic conformal algebra
$\SO(5,2)$.

%%%%%%%%%%%%%%%%%%%%%%%%%%%%%%%%%%%%%%%%%%%%%%%%%%%%%

\section{The current multiplet of Howe-Lindstr{\"o}m} \label{ss:howlin}

The supercurrent in 5 dimensions has been discussed before in the
literature~\cite{Howe:1981nz, Zucker:1999ej}. The authors
of~\cite{Howe:1981nz} found a $40+40$ current multiplet that couples to a
($32+32$) plus ($8+8$) reducible field multiplet. It turns out that also
the current multiplet itself is reducible. More precisely, the $40+40$
multiplet of~\cite{Howe:1981nz} reduces to our $32+32$ multiplet and an
additional $8+8$ multiplet.

The $40+40$ multiplet has all the currents of the $32+32$ multiplet in
table~\ref{tbl:currentmult}, except the current $b_{\m\n}$. In addition it
contains the currents which we present in table~\ref{tab:howlin}.
\begin{table}[htb]
\begin{center}
\begin{tabular}{||c|c|c|c|c||c|c||}
\hline \rule[-1mm]{0mm}{6mm} Current  & {$\SU(2)$} & $w$ & \# d.o.f. &
Expression & Redefined & $w$\\
\hline \rule[-1mm]{0mm}{6mm} $c$ & 1 & 2
 & 1 & $\s^2$ &$\widehat{c}$      &4\\
\rule[-1mm]{0mm}{6mm} $y^{(ij)}$  & 3 & 3
 & 3 & $\ft12\rmi \bar{\p}^i \p^j$ &$y^{ij}$&3\\
\rule[-1mm]{0mm}{6mm} $n_{[\m\n]}$  & 1 &  1 &  10 & $\s F_{\m\n} + \ft18
\rmi \bar{\p} \g_{\m\n} \p$& $b_{[\m\n]} + \widehat{a}_\m$ & 3\\
[1mm] \hline \rule[-1mm]{0mm}{6mm} $\chi^i$  & 2 & $5/2$ & 8 &
 $\s \p^i$ &$\widehat{\chi}^i$ &$7/2$ \\[1mm] \hline
\end{tabular}
\caption{\it The extra fields of the $40+40$ current multiplet. The column
``Redefined'' indicates the currents after the field redefinitions
(\ref{6comp}). The field $n_{\mu \nu}$ is not a Noether current, but after
 the field redefinition~(\ref{6comp}) it gives the Noether currents
$b_{[\mu \nu]}$ and $\widehat{a}_\m$. \label{tab:howlin}}
\end{center}
\end{table}

The multiplet of~\cite{Howe:1981nz} is generated by varying $\s^2$ under
supersymmetry until closure is reached and in this way it produces $40+40$
components.
In particular, we find that the transformations of the fields
in table~\ref{tab:howlin} are given by:
 \bea \d_Q c
&=& \rmi \bar{\e} \chi\, ,\nonumber\\
\d_Q \chi^i &=& - \ft14\rmi \slashed{\partial} c \e^i + \ft14 y^{ij} \e_j
+\ft14\rmi \g^\m v^{ij}_\m \e_j - \ft14 \g \cdot n \e^i\, , \nonumber\\
\d_Q y^{ij} &=& \rmi \bar{\e}^{(i} \rmi \g \cdot J^{j)} + 2 \bar{\e}^{(i}
\slashed{\partial}
\chi^{j)}\, ,\nonumber\\
\d_Q n_{\m\n} &=& \ft14 \bar{\e} \g_{\m\n\l} J^{\l} + \bar{\e}
\partial_{[\m} \g_{\n]} \chi\, . \label{extracurr}
 \eea
 In addition the transformation of
the supercurrent is also slightly changed w.r.t.\ (\ref{lintrcurr}) since
it does not contain $b_{\m\n}$ but $n_{\m\n}$:
 \be \d_Q J^i_\m = - \ft12\rmi \g^\n \th_{\n\m} \e^i - \rmi
\g_{[\l}\partial^\l v^{ij}_{\m]} \e_j - \ft12 a_\m \e^i + \ft14\rmi
\ve_{\m\n\l\r\s} \g^\n \partial^\l n^{\r\s} \e^i\, . \ee
Comparing this with~(\ref{lintrcurr}) we see that the relation between
$b_{\m\n}$ and $n_{\m\n}$ is given by
\be
b_{\m\n} = \ft12 \ve_{\m\n\l\r\s} \partial^\l n^{\r\s}\, .
 \ee
The $32+32$ components now transform only among themselves according
to~(\ref{lintrcurr}).  To demonstrate full reducibility of the $40+40$
multiplet we make the following field redefinitions:
 \bea
\widehat{c} &=& \th_\m{}^\m - \Box c\, , \nonumber\\
\widehat{\chi}^i
&=& \rmi \g \cdot J^i - 2 \rmi \slashed{\partial} \chi^i\, , \nonumber\\
\widehat{a}_\m &=& a_\m - 2 \partial^\n n_{\n\m}\, . \label{6comp}
 \eea
Together with the field $y^{ij}$ these fields form an $8+8$ multiplet,
see table~\ref{tab:howlin}, transforming only among themselves according
to \bea \d_Q \widehat{c}
&=& \ft12 \bar{\e} \slashed{\partial} \widehat{\chi}\,, \nn \\
\d_Q \widehat{\chi}^i &=& \ft12 \widehat{c} \e^i - \ft12 \rmi
\slashed{\partial} y^{ij} \e_j -
\ft12 \rmi \slashed{\widehat{a}} \e^i\, , \nn \\
\d_Q y^{ij}
&=& \rmi \bar{\e}^{(i} \widehat{\chi}^{j)}\, , \nn \\
\d_Q \widehat{a}_\m &=& -\ft12 \rmi \bar{\e} \g_{\m\n} \partial^\n
\widehat{\chi}\, . \label{88curr} \eea

As one would suspect from the field content, the $8+8$ current
multiplet~(\ref{88curr}) is conjugate to the off-shell vector
multiplet~(\ref{offshellvec}), in the same way as the $32+32$ current
multiplet~(\ref{lintrcurr}) is conjugate to the Dilaton Weyl
multiplet~(\ref{lintrDilW}). We expect that this multiplet can be used as
one of the compensating multiplets in the construction of the $48+48$
off-shell~\cite{Howe:1981ev,Zucker:1999ej} ($= 8+8$
on-shell~\cite{Cremmer:1980gs}) $D=5$ Poincar{\'e} multiplet.
%% For BibTeX
%\bibliography{refd5conf}

\begin{thebibliography}{10}

\bibitem{Salam:1989fm}
A.~Salam and E.~Sezgin, \emph{Supergravities in diverse dimensions. vol.
1,  2}, North-Holland 1989

\bibitem{Kaku:1978ea}
M.~Kaku and P.K. Townsend, \emph{Poincar{\'e} supergravity as broken
  superconformal gravity}, Phys. Lett. {\bf B76} (1978)
54.
%%CITATION = PHLTA,B76,54;%%.

\bibitem{Randall:1999vf}
L.~Randall and R.~Sundrum, \emph{An alternative to compactification},
Phys. Rev. Lett. {\bf 83} (1999) 4690--4693,
\href{http://www.arXiv.org/abs/hep-th/9906064}{{\tt hep-th/9906064}}.
%%CITATION = PRLTA,83,4690;%%.

\bibitem{Randall:1999ee}
L.~Randall and R.~Sundrum, \emph{A large mass hierarchy from a small extra
  dimension}, Phys. Rev. Lett. {\bf 83} (1999) 3370--3373,
\href{http://www.arXiv.org/abs/hep-ph/9905221}{{\tt hep-ph/9905221}}.
%%CITATION = PRLTA,83,3370;%%.

\bibitem{Nishimura:2000wj}
M.~Nishimura, \emph{Conformal supergravity from the AdS/CFT
correspondence},
  Nucl. Phys. {\bf B588} (2000) 471--482,
\href{http://www.arXiv.org/abs/hep-th/0004179}{{\tt hep-th/0004179}}.
%%CITATION = NUPHA,B588,471;%%.

\bibitem{D'Auria:2000ad}
R.~D'Auria, S.~Ferrara and S.~Vaula, \emph{Matter coupled $F(4)$
supergravity
  and the $AdS_6/CFT_5$ correspondence}, JHEP {\bf 10} (2000) 013,
\href{http://www.arXiv.org/abs/hep-th/0006107}{{\tt hep-th/0006107}}.
%%CITATION = JHEPA,0010,013;%%.

\bibitem{Balasubramanian:2000pq}
V.~Balasubramanian, E.~Gimon, D.~Minic and J.~Rahmfeld, \emph{Four
dimensional
  conformal supergravity from AdS space},
\href{http://www.arXiv.org/abs/hep-th/0007211}{{\tt hep-th/0007211}}.
%%CITATION = HEP-TH 0007211;%%.

\bibitem{Kallosh:2000tj}
R.~Kallosh and A.~Linde, \emph{Supersymmetry and the brane world}, JHEP
{\bf
  02} (2000) 005,
\href{http://www.arXiv.org/abs/hep-th/0001071}{{\tt hep-th/0001071}}.
%%CITATION = JHEPA,0002,005;%%.

\bibitem{Behrndt:2000tr}
K.~Behrndt and M.~Cveti\v c, \emph{Anti-de~Sitter vacua of gauged
supergravities
  with 8 supercharges}, Phys. Rev. {\bf D61} (2000) 101901,
\href{http://www.arXiv.org/abs/hep-th/0001159}{{\tt hep-th/0001159}}.
%%CITATION = PHRVA,D61,101901;%%.

\bibitem{Ceresole:2000jd}
A.~Ceresole and G.~Dall'Agata, \emph{General matter coupled ${\cal N} =
2$, $D
  = 5$ gauged supergravity}, Nucl. Phys. {\bf B585} (2000) 143--170,
\href{http://www.arXiv.org/abs/hep-th/0004111}{{\tt hep-th/0004111}}.
%%CITATION = NUPHA,B585,143;%%.

\bibitem{Gunaydin:1984bi}
M.~G{\"u}naydin, G.~Sierra and P.K. Townsend, \emph{The geometry of $N=2$
  Maxwell--Einstein supergravity and Jordan algebras}, Nucl. Phys. {\bf B242}
  (1984)
244.
%%CITATION = NUPHA,B242,244;%%.

\bibitem{Gunaydin:1985ak}
M.~G{\"u}naydin, G.~Sierra and P.K. Townsend, \emph{Gauging the $d = 5$
  Maxwell--Einstein supergravity theories: more on Jordan algebras}, Nucl.
  Phys. {\bf B253} (1985)
573.
%%CITATION = NUPHA,B253,573;%%.

\bibitem{Bergshoeff:2000zn}
E.~Bergshoeff, R.~Kallosh and A.~Van~Proeyen, \emph{Supersymmetry in
singular
  spaces}, JHEP {\bf 10} (2000) 033,
\href{http://www.arXiv.org/abs/hep-th/0007044}{{\tt hep-th/0007044}}.
%%CITATION = JHEPA,0010,033;%%.

\bibitem{Behrndt:2000km}
K.~Behrndt, C.~Herrmann, J.~Louis and S.~Thomas, \emph{Domain walls in
five
  dimensional supergravity with non- trivial hypermultiplets}, JHEP {\bf 01}
  (2000) 011,
\href{http://www.arXiv.org/abs/hep-th/0008112}{{\tt hep-th/0008112}}.
%%CITATION = JHEPA,0101,011;%%.

\bibitem{Behrndt:2000ph}
K.~Behrndt and M.~Cveti\v{c}, \emph{Gauging of $N = 2$ supergravity
  hypermultiplet and novel renormalization group flows}, to be published
  in Nucl. Phys. {\bf B},
\href{http://www.arXiv.org/abs/hep-th/0101007}{{\tt hep-th/0101007}}.
%%CITATION = HEP-TH 0101007;%%.

\bibitem{Behrndt:2001qa}
K.~Behrndt, S.~Gukov and M.~Shmakova, \emph{Domain walls, black holes, and
  supersymmetric quantum mechanics},
\href{http://www.arXiv.org/abs/hep-th/0101119}{{\tt hep-th/0101119}}.
%%CITATION = HEP-TH 0101119;%%.

\bibitem{Ceresole:2001wi}
A.~Ceresole, G.~Dall'Agata, R.~Kallosh and A.~Van~Proeyen,
  \emph{Hypermultiplets, domain walls and supersymmetric attractors},
  to be published in Phys. Rev. {\bf D},
\href{http://www.arXiv.org/abs/hep-th/0104056}{{\tt hep-th/0104056}}.
%%CITATION = HEP-TH 0104056;%%.

\bibitem{deWit:2001dj}
B.~de~Wit, M.~Ro\v{c}ek and S.~Vandoren, \emph{Hypermultiplets,
  hyperk{\"a}hler cones and quaternion-K{\"a}hler geometry}, JHEP {\bf 02}
  (2001) 039,
\href{http://www.arXiv.org/abs/hep-th/0101161}{{\tt hep-th/0101161}}.
%%CITATION = JHEPA,0102,039;%%.

\bibitem{Kaku:1977pa}
M.~Kaku, P.K. Townsend and P.~van Nieuwenhuizen, \emph{Gauge theory of the
  conformal and superconformal group}, Phys. Lett. {\bf B69} (1977)
304--308.
%%CITATION = PHLTA,B69,304;%%.

\bibitem{Kaku:1978nz}
M.~Kaku, P.K. Townsend and P.~van Nieuwenhuizen, \emph{Properties of
  conformal supergravity}, Phys. Rev. {\bf D17} (1978)
3179.
%%CITATION = PHRVA,D17,3179;%%.

\bibitem{Ferrara:1977ij}
S.~Ferrara, M.~Kaku, P.K. Townsend and P.~van Nieuwenhuizen, \emph{Unified
  field theories with $U(N)$ internal symmetries: gauging the superconformal
  group}, Nucl. Phys. {\bf B129} (1977)
125.
%%CITATION = NUPHA,B129,125;%%.

\bibitem{Bergshoeff:1986mz}
E.~Bergshoeff, E.~Sezgin and A.~Van~Proeyen, \emph{Superconformal tensor
  calculus and matter couplings in six dimensions}, Nucl. Phys. {\bf B264}
  (1986)
653.
%%CITATION = NUPHA,B264,653;%%.

\bibitem{Kugo:2000hn}
T.~Kugo and K.~Ohashi, \emph{Supergravity tensor calculus in $5D$ from
$6D$},
  Prog. Theor. Phys. {\bf 104} (2000) 835--865,
\href{http://www.arXiv.org/abs/hep-ph/0006231}{{\tt hep-ph/0006231}}.
%%CITATION = PTPKA,104,835;%%.

\bibitem{VanProeyen:1999ni}
A.~Van~Proeyen, \emph{Tools for supersymmetry},
  \href{http://www.arXiv.org/abs/hep-th/9910030}{{\tt hep-th/9910030}}.
in Annals of the University of Craiova, Physics AUC, Volume 9 (part I)
1999,
  pp.1--48.
%%CITATION = HEP-TH 9910030;%%.

\bibitem{Howe:1981nz}
P.~Howe and U.~Lindstrom, \emph{The supercurrent in five-dimensions},
Phys.
  Lett. {\bf B103} (1981)
422--426.
%%CITATION = PHLTA,B103,422;%%.

\bibitem{Ferrara:1975pz}
S.~Ferrara and B.~Zumino, \emph{Transformation properties of the
supercurrent},
  Nucl. Phys. {\bf B87} (1975)
207.
%%CITATION = NUPHA,B87,207;%%.

\bibitem{Sohnius:1979pk}
M.F. Sohnius, \emph{The multiplet of currents for $N=2$ extended
  supersymmetry}, Phys. Lett. {\bf B81} (1979)
8.
%%CITATION = PHLTA,B81,8;%%.

\bibitem{Bergshoeff:1981is}
E.~Bergshoeff, M.~de~Roo and B.~de~Wit, \emph{Extended conformal
  supergravity}, Nucl. Phys. {\bf B182} (1981)
173.
%%CITATION = NUPHA,B182,173;%%.

\bibitem{Bergshoeff:1999db}
E.~Bergshoeff, E.~Sezgin and A.~Van~Proeyen, \emph{(2,0) tensor
multiplets and
  conformal supergravity in $D = 6$}, Class. Quant. Grav. {\bf 16} (1999)
  3193--3206,
\href{http://www.arXiv.org/abs/hep-th/9904085}{{\tt hep-th/9904085}}.
%%CITATION = CQGRD,16,3193;%%.

\bibitem{Bergshoeff:1982av}
E.~Bergshoeff and M.~de~Roo, \emph{The supercurrent in ten-dimensions},
Phys.
  Lett. {\bf B112} (1982)
53.
%%CITATION = PHLTA,B112,53;%%.

\bibitem{deWit:1980ug}
B.~de~Wit, J.W. van Holten and A.~Van~Proeyen, \emph{Transformation rules
of
  $N=2$ supergravity multiplets}, Nucl. Phys. {\bf B167} (1980)
186.
%%CITATION = NUPHA,B167,186;%%.

\bibitem{Romans:1986tw}
L.~J. Romans, \emph{The $F(4)$ gauged supergravity in six-dimensions},
Nucl.
  Phys. {\bf B269} (1986)
691.
%%CITATION = NUPHA,B269,691;%%.

\bibitem{vanHolten:1982mx}
J.~W. van Holten and A.~Van~Proeyen, \emph{$N=1$ supersymmetry algebras
in $d = 2,3,4$ mod. 8}, J. Phys. {\bf A15}
(1982) 3763.
%%CITATION = JPAGB,A15,3763;%%.

\bibitem{D'Auria:2000ec}
R.~D'Auria, S.~Ferrara, M.A. Lledo and V.S. Varadarajan, \emph{Spinor
  algebras},
\href{http://www.arXiv.org/abs/hep-th/0010124}{{\tt hep-th/0010124}}.
%%CITATION = HEP-TH 0010124;%%.

\bibitem{Bergshoeff:1983az}
E.~Bergshoeff, M.~de~Roo and B.~de~Wit, \emph{Conformal supergravity in
ten
  dimensions}, Nucl. Phys. {\bf B217} (1983)
489.
%%CITATION = NUPHA,B217,489;%%.

\bibitem{Nahm:1978tg}
W.~Nahm, \emph{Supersymmetries and their representations}, Nucl. Phys.
{\bf
  B135} (1978)
149.
%%CITATION = NUPHA,B135,149;%%.

\bibitem{Zucker:1999ej}
M.~Zucker, \emph{Minimal off-shell supergravity in five dimensions}, Nucl.
  Phys. {\bf B570} (2000) 267--283,
\href{http://www.arXiv.org/abs/hep-th/9907082}{{\tt hep-th/9907082}}.
%%CITATION = NUPHA,B570,267;%%.

\bibitem{Howe:1981ev}
P.~Howe, \emph{Off-shell $N=2$ and $N=4$ supergravity in five-dimensions},
  CERN-TH-3181.

\bibitem{Cremmer:1980gs}
E.~Cremmer, \emph{Supergravities in 5 dimensions}, invited paper at the
  Nuffield Gravity Workshop, Cambridge, England, Jun 22--Jul 12, 1980.

\bibitem{Behrndt:1999kz}
K.~Behrndt and M.~Cveti\v{c}, \emph{Supersymmetric domain wall world from
$D =
  5$ simple gauged supergravity}, Phys. Lett. {\bf B475} (2000) 253--260,
\href{http://www.arXiv.org/abs/hep-th/9909058}{{\tt hep-th/9909058}}.
%%CITATION = PHLTA,B475,253;%%.

\bibitem{Fujita:2001kv}
T.~Fujita and K.~Ohashi, \emph{Superconformal tensor calculus in five
dimensions}, {\tt hep-th/0104130}.
%%CITATION = HEP-TH 0104130;%%


\end{thebibliography}
%\bibliographystyle{toine}
%\nocite{*}
%%%%%%%%%%%%
%
\providecommand{\href}[2]{#2}\begingroup\raggedright\endgroup

\end{document}